\pdfoutput=1
\documentclass{article}

\title{PACO: Global Signal Restoration via PAtch COnsensus 
\thanks{This paper is a significant extension of work previously published  in~\cite{paco-dct} and~\cite{paco-dct-ipol}. As such, some  sections are heavily based on the corresponding ones in the aforementioned works. Also, some results are repeated here for the sake of completeness.}
}
\author{Ignacio Ram\'{\i}rez~\thanks{Departamento de Procesamiento de Se\~{n}ales, Instituto de Ingenier\'{\i}a El\'{e}ctrica, Facultad de Ingenier\'{\i}a, Universidad de la Rep\'{u}blica, Uruguay.
(\texttt{nacho@fing.edu.uy}, \texttt{http://iie.fing.edu.uy/personal/nacho/})} }

\usepackage[utf8]{inputenc}
\usepackage{enumerate}
\usepackage{graphicx}
\usepackage[svgnames]{xcolor}
\usepackage{tikz}
\usepackage{multicol}

\usepackage{xspace}
\usepackage{amssymb}
\usepackage{amsmath}
\usepackage{amsfonts}
\usepackage{amsthm}

\newtheorem{theorem}{Theorem}

\newtheorem{corollary}[theorem]{Corollary}
\newtheorem{proposition}[theorem]{Proposition}

\usepackage{url} 
\usepackage{balance}
\usepackage[margin=2cm]{geometry}
\usepackage{algorithm}[0.1]
\usepackage{algorithmic}
\usepackage[capitalize,nameinlink]{cleveref}[0.19]
\crefname{section}{section}{sections}
\crefname{subsection}{subsection}{subsections}
\Crefname{section}{Section}{Sections}
\Crefname{subsection}{Subsection}{Subsections}

\Crefname{figure}{Figure}{Figures}

\crefformat{equation}{\textup{#2(#1)#3}}
\crefrangeformat{equation}{\textup{#3(#1)#4--#5(#2)#6}}
\crefmultiformat{equation}{\textup{#2(#1)#3}}{ and \textup{#2(#1)#3}}
{, \textup{#2(#1)#3}}{, and \textup{#2(#1)#3}}
\crefrangemultiformat{equation}{\textup{#3(#1)#4--#5(#2)#6}}%
{ and \textup{#3(#1)#4--#5(#2)#6}}{, \textup{#3(#1)#4--#5(#2)#6}}{, and \textup{#3(#1)#4--#5(#2)#6}}

\Crefformat{equation}{#2Equation~\textup{(#1)}#3}
\Crefrangeformat{equation}{Equations~\textup{#3(#1)#4--#5(#2)#6}}
\Crefmultiformat{equation}{Equations~\textup{#2(#1)#3}}{ and \textup{#2(#1)#3}}
{, \textup{#2(#1)#3}}{, and \textup{#2(#1)#3}}
\Crefrangemultiformat{equation}{Equations~\textup{#3(#1)#4--#5(#2)#6}}%
{ and \textup{#3(#1)#4--#5(#2)#6}}{, \textup{#3(#1)#4--#5(#2)#6}}{, and \textup{#3(#1)#4--#5(#2)#6}}

\crefdefaultlabelformat{#2\textup{#1}#3}

\definecolor{NewTextFG}{rgb}{0.0,0.1,0.2}

\newcommand{\best}[1]{\textbf{\color{blue}#1}\xspace}

\newcommand{\ok}[1]{\marginpar{\framebox{\large\color{blue}\checkmark}}}

\newcommand{\iter}[1]{^{(#1)}}
\newcommand{\BlackBox}{\rule{1.5ex}{1.5ex}}  

\def\reals{\ensuremath{\mathbb{R}}}

\renewcommand{\vec}[1]{\ensuremath{\mathbf{\MakeLowercase{#1}}}}
\newcommand{\mat}[1]{\ensuremath{\mathbf{\MakeUppercase{#1}}}}

\newcommand{\st}{\ensuremath{\mathrm{s.t.}}}

\newcommand{\fun}[1]{\mathrm{#1}}

\newcommand{\setdef}[1]{\ensuremath{\left\{#1\right\}}}

\def\transp{^\intercal}

\def\Gaussian{\ensuremath{\mathcal{N}}}

\def\inv{^{-1}}

\def\vecop{\mathrm{vec}}

\def\ident{\mat{I}}

\def\spX{\mathbb{X}}
\def\spY{\mathbb{Y}}
\def\spG{\mathbb{G}}
\def\spC{\mathbb{C}}

\def\oS{\mathcal{S}}
\def\ost{\mathcal{T}}
\def\vecop{\mathrm{vec}}
\def\matop{\mathrm{mat}}
\def\projop{\mathrm{\Pi}}
\newcommand{\prox}[2]{\ensuremath{\mathrm{prox}_{#1}\left(#2\right)}}
\def\mR{\mat{R}}
\def\oR{\mathcal{R}}
\def\va{\vec{a}}

\def\ve{\vec{e}}
\def\vu{\vec{u}}

\def\vx{\vec{x}}

\def\vy{\vec{y}}
\def\vz{\vec{z}}

\def\mA{\mat{A}}
\def\mB{\mat{B}}

\def\mD{\mat{D}}

\def\mI{\mat{I}}

\def\mU{\mat{U}}

\def\mY{\mat{Y}}
\def\mZ{\mat{Z}}

\def\hx{\hat{x}}

\def\hz{\hat{z}}

\def\hmA{\hat{\mA}}
\def\hmY{\hat{\mY}}
\def\hva{\hat{\va}}

\def\hvx{\hat{\vx}}
\def\hvy{\hat{\vy}}
\def\hvz{\hat{\vz}}

\def\tvx{\tilde{\vx}}

\newcommand{\acro}[1]{\textsc{\MakeLowercase{#1}}\xspace}
\def\admm{\acro{admm}}
\def\ladmm{\acro{ladmm}}
\def\paco{\acro{paco}}
\def\dct{\acro{dct}}
\def\rmse{\acro{rmse}}

\begin{document}
%
\maketitle
\begin{abstract}
Many signal processing algorithms break the target signal into overlapping segments (also called windows, or patches), process them separately, and then stitch them back into place to produce a unified output. At the overlaps, the final value of those samples that are estimated more than once needs to be decided in some way. Averaging, the simplest approach, often leads to unsatisfactory results.  Significant work has been devoted to this issue in recent years. Several works explore the idea of a weighted average of the overlapped patches and/or pixels; others promote agreement (consensus) between the patches at their intersections. Agreement can be either encouraged or imposed as a hard constraint. This work develops on the latter case. The result is a variational signal processing framework, named PACO, which features a number of appealing theoretical and practical properties. The PACO framework consists of a variational formulation that fits a wide variety of problems, and a  general \admm-based algorithm for minimizing the resulting energies. As a byproduct, we show that the consensus step of the algorithm, which is the main bottleneck of similar methods, can be solved efficiently and easily for any arbitrary patch decomposition scheme. 
We demonstrate the flexibility and power of PACO on three different problems: image inpainting (which we have already covered in previous works), image denoising, and contrast enhancement, using different cost functions including  Laplacian and Gaussian Mixture Models. 
\end{abstract}
%

\textbf{Key words.} patch-based methods, global-local methods, patch consensus, inpainting, denoising, contrast enhancement.\\

\textbf{AMS subject classifications.} 49M30, 49M37, 65K05, 65K10, 90C25, 90C30.\\

\section{Introduction}
%
\label{sec:intro}

We refer to \emph{patch restoration methods} to a family of techniques where the target signal is first broken down into smaller, possibly overlapping patches of some size and shape, some restoration method is applied to each patch separately, and the restored patches are stitched back together into their corresponding place to produce the complete restored signal (sometimes referred to as the \emph{global} signal). 
This is a common technique, with many examples in audio (e.g.~\cite{audio-ms,audio4,audio3,audio-stft}) and image processing (see e.g.,~\cite{bm3d, elad06, patch-wiener, sparse-model-online, sparse-model-color, sparse-model-ms,  patch-ms}). A recent review on patch-based restoration methods for image processing can be found in~\cite{patch-review}.  Note that we are not considering  methods such as \cite{nlm, idude} (and their many variants) which estimate each sample separately using the surrounding patches as context, but only those that process the whole patch as a sub-signal.

When overlapping occurs, the final value of those samples that are estimated more than once must be resolved in some way. A traditional approach in audio processing methods is to apply a windowing function to the patches (e.g. Hanning); this procedure is backed by theoretical results related to the violation of frequency-based hypothesis during processing, but is widely applied to other methods as well. On the other hand, many recent, very successful patch-based methods do away with the overlapping issue by simply averaging the overlapped patches. This may incidentally help in covering up artifacts in the restoration process, but more often will simply smooth out the resulting global estimate (see e.g.\cite{elad06}). To overcome such limitations via better stitching strategies has been the subject of several works over the last decade: some of them~\cite{patch-weighting3, patch-weighting1,patch-weighting2,patch-weighting4,patch-weighting5} are based on assigning weights to the different patches when averaging; others such as ~\cite{arias, patch-weighting-pixel} assign a different weight to each pixel. We refer to these as \emph{weighted averaging methods}. 

An alternative to weighted averaging, which sidesteps the arbitrariness of defining appropriate weights, is to promote solutions where patches coincide at their intersection. This idea appears under the keyword \emph{patch disagreement} in~\cite{patch-disagreement}; there, an iterative method is used to give more weight to the discrepancies between the estimated patches, with the aim of reducing such disagreement in the long run. 

The problem of patch aggregation has also be studied from a probabilistic viewpoint in works such as~\cite{epll} and~\cite{patch-aggregation}.
The work ~\cite{epll} is based on the observation that, after stitching, the averaged patches may no longer follow the prior from which they were modeled. The authors of~\cite{epll} seek a remedy to this by augmenting the energy to be minimized with a quadratic term that penalizes the disagreement between the patches before and after stitching.

In~\cite{patch-aggregation}, instead of directly averaging  the patch estimates, the authors produce a unique probability model over the whole image by ``stitching'' the probability models of each independent patch via a well-defined fusion mechanism. Besides its own value as an original and powerful concept, the authors of~\cite{patch-aggregation} derive a new interpretation on the aforementioned work~\cite{epll} as a particular fusion mechanism.

Back to the problem of averaging patch estimations, it was observed in ~\cite{figueiredo} that the agreement imposed in~\cite{epll} would become a consensus constraint on the patches as the quadratic penalty coefficient grew to infinity. They then draw on this idea to propose a general consensus-based \acro{map} denoising algorithm and a general \admm-based method for solving it. In this sense, the method proposed in~\cite{figueiredo} can be seen as a particular case of our proposed framework. 
Furthermore, the work~\cite{figueiredo} analyzes the computational cost and bottlenecks of their proposed \admm method to some degree. However, they fall short of fully exploiting the structure of the problem, leading to an overly conservative upper bound on the overall computational cost of the method. As we show below, this cost can be significantly reduced.


\subsection{Contributions}

This work proposes a general variational formulation for patch-based restoration methods, under a strict patch consensus constraint which can be applied to a wide family of problems, linear and non-linear, convex and non-convex, sparse or dense, to signals of any dimension. Below we summarize the contributions of our work.

\paragraph{Patch Consensus Framework} We provide a formal characterization of the Patch Consensus set, its geometrical properties, and its relationship with the space of signals. We then develop a general formulation for patch restoration problems under the \paco constraint which can be applied to any method that can be written as the minimization of a cost function of the estimated patches and/or signal; this can be seen as a generalization of the one proposed in~\cite{figueiredo}. 

\paragraph{Optimization} We propose a method for solving the aforementioned family of problems based on a splitting strategy and the standard \admm~\cite{admm} algorithm. While our algorithm bears strong similarities to that proposed in~\cite{figueiredo}, the latter is limited to the denoising case. Also, by retaining the patch extraction and stitching operations in matrix form, it does not fully exploit the geometry of the patch consensus space, which leads to a significantly higher computational cost. 

In contrast, our \admm-based  algorithm can be readily applied to any variational patch-based restoration method. We also develop the special case of orthogonal linear patch models, and a Linearized \acro{admm} algorithm~(a.k.a. Uzawa's Method~\cite{uzawa}) for non-orthogonal linear patch models, which has the same convergence guarantees as \admm while being computationally feasible even for very large signals. Both methods require significantly less computational resources than previously-proposed patch-consensus algorithms, and inherit the convergence properties of the \admm method.

\paragraph{Stitching trick} As described earlier, the bottleneck of methods such as \cite{epll,figueiredo,global-local} lies in the  ``patch agreement step'', which might not be practical even for small problem sizes. In this work we show that the projection operator onto the consensus constraint set boils down to a simple composition of the patch stitching and extraction operations. This not only avoids the computation and/or storage of projection matrices, but is also very easy to implement, and generalizes to any arbitrary patch decomposition scheme including irregular grids, different patch sizes, etc., on signals of any dimension.

\paragraph{Global constraints} As mentioned above, the \paco constraint guarantees a one-to-one correspondence between patch space (sometimes called ``local'' space) and global signal space; this allows formulations where arbitrary constraints are imposed on the global signal rather than on the patches. This is a crucial advantage of imposing strict consensus rather than promoting it in a lax way; we show this feature in the missing samples estimation method described \cref{sec:inpainting}. Furthermore, for the types of patch extraction operations used in typical patch-based applications, the mapping between signal space and patch space can be considered to be an isometry. This allows for orthogonal projections on constraint sets to be performed on either space.

\paragraph{Applications and variants} We show the broad applicability and the potential of \paco on three very different settings: inpainting, denoising, and contrast enhancement. Each problem is discussed in a self-contained section where we provide a brief introduction to the problem and derive the corresponding optimization algorithms. Note that the inpainting problem has already been covered in detail in~\cite{paco-dct,paco-dct-ipol}; we repeat the main results here for completeness.

In order to further demonstrate the flexibility of the framework, different patch priors are explored in each case: we use Laplacian prior on \dct coefficients for inpainting, and a Gaussian Mixture Model~\cite{gmm} for denoising.
The \dct-based inpainting case has already been covered in detail in~\cite{paco-dct} and \cite{paco-dct-ipol}. For the other settings we provide preliminary results; a complete treatment of these methods, as well as experiments using other priors, will be published elsewhere.

\subsection{Organization}

The rest of the document is laid out as follows. We begin with an introduction to the general problem of patch-based signal processing in \cref{sec:patch-methods}. %
\Cref{sec:paco} provides a formal definition of the \paco family of problems, as well as a general solution of the problem which serves as a basis for all the algorithms developed later. %
The next three sections provide self-contained treatments of different signal processing problems using \acro{paco}:
our previous work on inpainting is revisited in \cref{sec:inpainting}, \cref{sec:denoising} deals with denoising, and \cref{sec:contrast} introduces \paco for contrast enhancement. We conclude our treatment in \cref{sec:conclusion}.

\section{Patch-based methods}
\label{sec:patch-methods}

The methods that we are interested in this paper involve three stages that may be repeated until some convergence criterion is met: \emph{patch extraction}, \emph{patch estimation}, and \emph{patch stitching}. Recall that we are interested in methods that estimate \emph{whole} patches rather than a single pixel (e.g., the center) such as in \cite{nlm}. We will now illustrate these concepts on a simple one-dimensional case. 

\subsection{Patch extraction}
\label{sec:patch-extraction}

\begin{figure}
\centering%
\begin{tikzpicture}[xscale=0.4,yscale=0.2]
\draw [orange,,<-] (7.5,7.5) -- (1.5,1.5);
\draw [orange,<-] (9.5,7.5) -- (3.5,1.5);
\draw [orange,<-] (11.5,7.5) -- (5.5,1.5);
\draw [orange,<-] (13.5,7.5) -- (7.5,1.5);
\draw [orange,<-] (13.5,5.5) -- (9.5,1.5);
\draw [orange,<-] (13.5,3.5) -- (11.5,1.5);

\draw (7,3) -- (7,9); 
\draw (9,3) -- (9,9); 
\draw (11,3) -- (11,9);
\draw (13,3) -- (13,9);
\draw (15,3) -- (15,9);
\draw (7,3) -- (15,3);
\draw (7,5) -- (15,5);
\draw (7,7) -- (15,7);
\draw (7,9) -- (15,9);
\node at (8,8) {${x}_{1}$};
\node at (10,8) {${x}_{2}$};
\node at (12,8) {${x}_{3}$};
\node at (14,8) {${x}_{4}$};
\node at (8,6) {${x}_{2}$};
\node at (10,6) {${x}_{3}$};
\node at (12,6) {${x}_{4}$};
\node at (14,6) {${x}_{5}$};
\node at (8,4) {${x}_{3}$};
\node at (10,4) {${x}_{4}$};
\node at (12,4) {${x}_{5}$};
\node at (14,4) {${x}_{6}$};
\draw (0,0) -- (12,0);
\draw (0,2) -- (12,2);
\draw (0,0) -- (0,2);
\draw (2,0) -- (2,2);
\draw (4,0) -- (4,2);
\draw (6,0) -- (6,2);
\draw (8,0) -- (8,2);
\draw (10,0) -- (10,2);
\draw (12,0) -- (12,2);
\node at (1,1) {${x}_{1}$};
\node at (3,1) {${x}_{2}$};
\node at (5,1) {${x}_{3}$};
\node at (7,1) {${x}_{4}$};
\node at (9,1) {${x}_{5}$};
\node at (11,1) {${x}_{6}$};

\node at (1,3) {$\mathbf{x}$};
\node at (11,10) {$\mathbf{Y}$};
\end{tikzpicture}
\caption{\label{fig:patch-extraction} Patch extraction operator $\mY=\oR(\vx)$ for a signal $\vx$ of length $N=6$ and patches of size $m=3$; patches are arranged as columns on an $m{\times}n$ matrix $\mY$ where the column $\vec{y}_k$ contains the patch starting at offset $k$ in $\vx$.}
\end{figure}

\begin{figure}
\centering%
\begin{tikzpicture}[xscale=0.4,yscale=0.2]
\draw [cyan,->] (8,8) -- (1.5,1.5);
\draw [cyan,->] (10,8) -- (3.5,1.5);
\draw [cyan,->] (12,8) -- (5.5,1.5);
\draw [cyan,->] (14,8) -- (7.5,1.5);
\draw [cyan,->] (14,6) -- (9.5,1.5);
\draw [cyan,->] (14,4) -- (11.5,1.5);
\draw (7,3) -- (7,9); 
\draw (9,3) -- (9,9); 
\draw (11,3) -- (11,9);
\draw (13,3) -- (13,9);
\draw (15,3) -- (15,9);
\draw (7,3) -- (15,3);
\draw (7,5) -- (15,5);
\draw (7,7) -- (15,7);
\draw (7,9) -- (15,9);
\node at (8,8) {$\hat{y}_{11}$};
\node at (10,8) {$\hat{y}_{12}$};
\node at (12,8) {$\hat{y}_{13}$};
\node at (14,8) {$\hat{y}_{14}$};
\node at (8,6) {$\hat{y}_{21}$};
\node at (10,6) {$\hat{y}_{22}$};
\node at (12,6) {$\hat{y}_{23}$};
\node at (14,6) {$\hat{y}_{24}$};
\node at (8,4) {$\hat{y}_{31}$};
\node at (10,4) {$\hat{y}_{32}$};
\node at (12,4) {$\hat{y}_{33}$};
\node at (14,4) {$\hat{y}_{34}$};
\draw (0,0) -- (12,0);
\draw (0,2) -- (12,2);
\draw (0,0) -- (0,2);
\draw (2,0) -- (2,2);
\draw (4,0) -- (4,2);
\draw (6,0) -- (6,2);
\draw (8,0) -- (8,2);
\draw (10,0) -- (10,2);
\draw (12,0) -- (12,2);
\node at (1,1) {$\hat{x}_{1}$};
\node at (3,1) {$\hat{x}_{2}$};
\node at (5,1) {$\hat{x}_{3}$};
\node at (7,1) {$\hat{x}_{4}$};
\node at (9,1) {$\hat{x}_{5}$};
\node at (11,1) {$\hat{x}_{6}$};
\node at (1,3) {$\hat{\mathbf{x}}$};
\node at (11,10) {$\hat{\mathbf{Y}}$};
\end{tikzpicture}
\caption{\label{fig:patch-stitching}Average patch stitching operator $\hvx=\oS(\hmY)$. Each sample $\hat{x}_j$ is the average of all its estimates in $\hmY$. In our example, this corresponds to the average of each anti-diagonal of $\hmY$.}
\end{figure}

Given a 1D signal $\vx=(x_1,\ldots,x_N) \in \spX=\reals^N$  of length $N$, the patch extraction is a linear operator $\oR$ determined by a \emph{patch extraction matrix} $\mR\transp = [\;\mR_1\transp\;|\;\mR_2\transp\;|\;\cdots\;|\;\mR_n\transp\;],$ where each sub-matrix $\mR_j, j=1,\ldots,n$ copies (extracts) samples from $\vx$ into a sub-vector $\vy_j$, which we call a patch. Which samples, how many samples, and how many copies of each sample in $\vx$ are extracted onto $\vy_j$ is arbitrary and may depend on $j$, so that the dimension of each patch is some  $m_j \in \mathbb{N}$. The result of the extraction operator is a \emph{patches vector} $\vy\transp=[\vy_1\transp\;|\;\vy_2\transp\;|\;\ldots\;|\;\vy_n\transp] \in \reals^{\sum_{j=1}^{n}{m_j}}$. The space $\spY=\reals^{\sum_{j=1}^{n}{m_j}}$ is called the \emph{patches space}.
In the common case where $m_j=m$ for all $j$, it may be convenient to rearrange the patches vector $\vy \in \reals^{mn}$ as a matrix $\mY \in \reals^{m\times{n}}$, where the patches $\vy_j$ are arranged as columns; many formulations are presented in this way hereafter. Formally, however, we will always refer to the extracted patches as a vector belonging to the \emph{patch space} $\spY$.

In the example of \cref{fig:patch-extraction} we have a 1D signal of length $N=6$, patches of size $m=3$ and  $\mR_j=\left[\ve^{j}\;|\;\ve^{j+1}| \ve^{j+2} \right]$ $j=1,2,3,4$. 
More generally, we will have %
$\mR_j\transp=\left[ \ve^{1} \;|\; \ve^{1+s} \;|\; \ve^{1+2s} \;|\; \ldots \;\right]$ where $s > 1$ is called the \emph{stride}, that is, the separation in space between one patch an its neighbor. The case $s=1$ depicted in \cref{fig:patch-extraction} yields \emph{fully overlapped} patches; the extreme case $s=3$ yields non-overlapping patches. 
The preceding discussion generalizes to signals $\vx$ of any dimension by \emph{vectorizing} $\vx$ prior to extraction: $\vy_j = \mR_j \vecop(\vx)$. Here $\vecop(\cdot)$ arranges its argument into a vector by traversing its elements in some predefined order. In our case, for $2D$ signals of size $M{\times}N$, we follow a row-major ordering, that is, the output vector is the concatenation of the $M$ input rows.  The  inverse of $\vecop(\cdot)$ is called the \emph{matrification} operator and is denoted by $\matop(\cdot)$.

\paragraph{Geometry of the problem}
The operator $\oR$ is linear and injective. Thus, its image $\spC$ is a subspace of the patches space $\spY$. If $\oR$ extracts all the samples from $\vx$ then it also defines a linear bijection between $\spX$ and $\spC$. As defined, the elements of $\spC$ which are mapped from the same sample in $\vx$ are all equal; for reasons that will become clear soon, we say that these elements are \emph{in consensus}. Thus we call $\spC$ the \emph{consensus set}. 

The sets $\spX$ and $\spC$ are, however, not isometric. It is easy to verify that this can only happen if $\spC$ contains exactly the same number of copies of each sample in $\spX$. Most common patch extraction schemes used in the literature do not comply with this, as samples in the borders usually belong to fewer patches than those in the center.
However, for most signals and, in particular images, of practical interest, the border samples are few  and thus the mappings $\spX$ and $\spC$ can be considered isometric for most practical purposes. This will prove important, as it allows to perform orthogonal projections onto convex sets defined in $\spC$ in the (usually much smaller) space $\spX$.

\subsection{Patch averaging and the stitching trick}

In a signal restoration setting one does not observe the \emph{true} or \emph{clean} signal $\tvx$, but a distorted version $\vx$ (which usually has the same size and dynamic range as $\tvx$), and the task is to infer $\tvx$ from $\vx$. We denote the result of this inference as $\hvx$, and call it the \emph{restored} or \emph{estimated} signal indistinctly.
The patches vector extracted from $\vx$ is correspondingly denoted by $\vy$. The idea is to estimate the signal $\tvx$ by recomposing it from an estimate $\hvy$, which is itself a function of $\vy$. In general, however, $\vy\notin\spC$. 

By \emph{stitching} we refer to the operation that maps a patches vector $\hvy \in \spY$ onto an estimated signal $\hvx$. In general, this operator is not unique. A common choice is to seek for the signal estimate $\hvx$ whose patches, when extracted, are as close as possible to the individually estimated ones in $\hvy$. If we measure this distance using $\ell_2$ norm, we obtain the classical \emph{least squares} estimator: 
\begin{equation}
\hvx_{ls} = \arg\min_\vx \|\mR\vx - \hvy\|_2 = \left(\sum_j \mR_j\transp\mR_j\right)^{-1} \sum_j \mR_j\transp \hvy_j = (\mR\transp\mR)\inv\mR\transp\hvy.
  \label{eq:ops} 
\end{equation}
 
\Cref{eq:ops} is interpreted as follows: it is easy to verify that $\mR_j\transp\hvy_j$ puts back the patch $\hvy_j$ in its corresponding place in $\hvx$; thus, $\mR\transp\hvy$ produces a vector of length $N$ where the $i$-th element contains the sum of all the elements of $\hvy$ that are mapped there by one or more $\mR_j$.  
On the other hand $\mR_j\transp{\mR_j}$ is a diagonal matrix with $m$ ones, and so $\mR\transp\mR=\sum_j \mR_j\transp{\mR_j}$ is a diagonal matrix whose $(i,i)$-th element counts how many $\mR_j$s have mapped some element of $\hvy$ to  the sample $i$. Thus, the $i$-th entry of the right hand side of~\cref{eq:ops} is the average of all the different estimates of the $i$-th sample in $\hvy$. 
We call the operator defined in \cref{eq:ops} as the \emph{average stitching operator} and denote it by $\oS:\spY \rightarrow \reals^N$, $\hvx=\oS(\hvy)$.
The extraction operator $\oR$ is given by $\vy = \oR(\vx) = \mR \vx$. If we compose this with the stitching operator $\oS$ defined earlier we obtain 
\begin{equation}
  \oR[ \oS (\hvy) ] = \mR(\mR\transp\mR)\inv\mR\transp \hvy = \projop_\spC(\hvy),
  \label{eq:proj-op}
\end{equation} 
where $\projop_{\spC}(\hvy)$ is the orthogonal projection of $\hvy \in \spY$ onto $\spC$. Thus, we conclude that the projector operator~\cref{eq:proj-op} from $\spY$ onto $\spC$ can be written as the composition of the extraction operator with its corresponding average stitching operator. Although~\cref{eq:proj-op} is a well known result, to the best of our knowledge, its straightforward implementation as the aforementioned composition has not been fully exploited in the literature; this is why we call it a \emph{trick}. Rather, works such as \cite{global-local,figueiredo} compute the orthogonal projection matrix $\mR(\mR\transp\mR)\inv\mR\transp$ explicitly, something that quickly becomes unwieldy as the signal size increases even if the structure of such matrix is exploited (e.g., block sparsity). 
This is further complicated if, instead of the patches, the stitching is written in terms of some affine transformation of the patches such as $\vy_j=\mD\va_j + b$, such as in \cite{global-local,figueiredo}.
The \emph{stitching trick} will only work directly on patches, or some orthogonal transformation of them. Later we will see how to deal with the affine case. What is also remarkable is that the procedure does works for any arbitrary extraction operator $\oR$. In particular, it does not require patches to be of the same shape and/or number of samples.

\section{PACO: The PAtch COnsensus problem}
\label{sec:paco}
\subsection{General consensus problems}
\label{sec:consensus}

This subsection is based on the monograph~\cite[Chapter 5]{proximal}.

In the context of parallel and distributed optimization, a common problem is to minimize a function $f(\zeta) = \sum_{j=1}^{r} f_j(\zeta)$, where $f_j(\zeta)$ are functions that can only be evaluated \emph{locally}, say at some node $j$ on a spatially distributed network with $r$ nodes, but which all depend on the same \emph{global} variable $\zeta \in \reals^m$. 
One strategy to solve this problem in a decentralized way is for each node to have its own \emph{local copy} of $\zeta$, $\zeta^j$, which it then optimizes in terms of its local function $f_(\zeta^j)$. In the end, however, we want all the $\zeta^j$ to be equal. For this we define the \emph{consensus set} as 
$\spG = \setdef{(\zeta^1,\zeta^2,\ldots,\zeta^r) \in \reals^{m{\times}r}: \zeta^1=\zeta^2=\ldots=\zeta^r}$ 
and constrain the final solution to fall within $\spG$. This gives rise to the following \emph{global consensus problem}:
\begin{equation*}
  \min_{(\zeta^1,\zeta^2,\ldots,\zeta^r) \in \spG} \sum_{j=1}^{r}f_j(\zeta^j).
\end{equation*}
Clearly, $\spG$ is a linear subspace of $\reals^{m{\times}r}$ (e.g., the line $x=y$ in $\reals^{2}$). %
Let $\projop_\spG(\cdot)$ denote the orthogonal projection operator from $\reals^{m{\times}r}$ onto $\spG$. It is straightforward to verify that,
\begin{equation*}
  \projop_\spG{(\zeta^1,\zeta^2,\ldots,\zeta^r)} = (\bar{\zeta},\bar{\zeta},\ldots,\bar{\zeta}),\quad \bar{\zeta} = \frac{1}{r}\sum_j \zeta^j.
\end{equation*}
Let $[m]$ be a shorthand for the set $\{1,2,\ldots,\}$. Consider the more general case where  each local copy $\zeta^j$ has an index subset $c_j \subseteq [m]$ associated to it, and consensus with any other copy $\zeta^{j'}$ is required only for those entries indexed by $c_j \cap c_{j'}$. The  \emph{partial consensus set} $\spG'$ is defined as follows: 
\begin{equation}
  \spG' = \setdef{(\zeta^1,\zeta^2,\ldots,\zeta^r): \zeta^j_i = \zeta^{j'}_i, \forall\,i \in c_{j} \cap c_{j'}, \forall\,j,j' \in [n], j \neq j'}.
  \label{eq:consensus-set}
\end{equation}
The corresponding \emph{partial consensus problem} is  given by:
\begin{equation}
  \min_{(\zeta^1,\zeta^2,\ldots,\zeta^r) \in \spG'} \sum_{j=1}^{r}f_j(\zeta^j).
  \label{eq:general-consensus-problem}
\end{equation}
It is easy to show that the projection onto $\spG'$ is given by
\begin{equation}
  \projop_{\spG}(\zeta^j_i) = \bar{\zeta}_i: \bar{\zeta}_i = \frac{\sum_{j: i \in c_j} \zeta^j_i}{\sum_{j: i \in c_j} 1 }.
\label{eq:general-consensus}
\end{equation}
In words, after projection, the $i$-th element of all vectors that include $i$ in their corresponding set $c_j$ will be the average of the values of the $i$-th elements on those vectors before projection.

\subsection{Patch consensus problem}
\label{sec:paco-problem}

Given a patch extraction operator $\oR: \spX \rightarrow \spY$ with its 
corresponding consensus set $\spC$, a cost function $f(\vy):\spY \rightarrow \reals$, and an optional constraint set $\Omega \subseteq \spY$, the \paco problem is stated as follows:
\begin{equation}
  \hvy = \arg\min_{\vz} f(\vz) \quad\st\quad \vz \in \Omega \cap \spC.
  \label{eq:paco-gen}
  \end{equation}

We begin by showing that, when $\Omega = \spY$, \cref{eq:paco-gen} is a especial case of \cref{eq:general-consensus}. We will  treat the general case $\Omega \neq \spC$ later. At this point, we shall remind the reader that $\spC$ is isomorphic to $\spX$, so that the constraint set $\Omega$ can be defined in either $\spX$ or $\spC$, or a combination of both. For example, in order to clip the estimated samples to the range  $[0-255]$ we could define  $\Omega =\{\vz \in \spY: \oS(\vz) \in [0,255]^N \}$. 

\begin{proposition}
The patch consensus set $\spC$ is a particular case of the general consensus set $\spG'$ and so  the \paco  problem is of the form~\cref{eq:general-consensus-problem}.
\begin{proof}
  It suffices to define $\zeta_j = \mR_j\transp \vy_j$ and $c_j = \fun{nonzero}(\fun{diag}(\mR_j\transp{\mR_j}))$. Here $\fun{diag}(\cdot)$ returns the diagonal of an $N{\times}N$ square matrix as a vector of length $N$, and $\fun{nonzero}(\cdot)$ returns the indices of the nonzero elements in its argument. Note that, as defined, $c_j$ contains the indices of the samples filled in by $\vy_j$. Thus, the requirement that $\vy_j$ and $\vy_{j'}$ should coincide at $c_j \cap c_{j'}$ matches the definition given before for the consensus set in \cref{eq:consensus-set} for this particular choice of $\zeta_j$ and $c_j$.
  Now, given $f(\vy)=\sum_j f_j(\vy_j)$ in \cref{eq:paco-gen}, we define $f'_j(\zeta_j) = f_j(\mR\zeta_j)=f_j(\vy_j)$ and plug it into \cref{eq:general-consensus-problem}.
\end{proof}
\end{proposition}

\begin{corollary}
  The orthogonal projection onto the patch consensus set $\spC$ is given by the \emph{stitching trick}, $\projop_\spC(\hvy) = \oR[\oS(\hvy)]$, as in \cref{eq:proj-op}. 
\begin{proof}
  We have established that the patch consensus set $\spC=\spG'$, and that \cref{eq:proj-op} gives $\projop_\spC(\hvy)$. Thus, $\projop_\spC({\hvy}) = \projop_\spC({\hvy}) = \oR[ \oS (\hvy)]$.
\end{proof}
\end{corollary}

\subsection{Optimization}
\label{sec:paco-optimization}

We define the \emph{convex indicator function} associated to the \paco constraint set $\spC \cap \Omega$, $g(\cdot):\spY \rightarrow \reals\cup\{+\infty\}$,
\begin{equation*}
  g(\xi) = 
  \left\{
    \begin{array}{lcl}
      0 &,& \xi \in \spC \cap \Omega \\
      +\infty&,& \xi \notin \spC \cap \Omega
  \end{array}
  \right.
\end{equation*}
This allows us to transform the \paco problem into an unconstrained one,
\begin{equation}
\hvy = \arg\min_{\xi} f(\xi) + g(\xi).\label{eq:paco-gen2}
\end{equation}
The problem \cref{eq:paco-gen2} will be convex if the function $f(\cdot)$ and the set $\Omega$ are also convex. This allows for convex restoration problems to be defined under consensus constraints. This being said, the \paco framework~\cref{eq:paco-gen2} is not limited to convex problems.

The following method is based on the \emph{proximal operator form}~\cite{proximal} of the popular \emph{Alternating Directions Method of Multipliers} (\admm)~\cite{admm}. The proximal operator of a function $f(\cdot)$ with parameter $\lambda$ is given by,
\begin{equation}
  \prox{\lambda f}{y} := \arg\min_\xi f(\xi) + \frac{1}{2\lambda}\|\xi-y\|^2.
\end{equation}
The proximal operator has many interesting properties (see~\cite{proximal} and references therein). Among them, if $g(\cdot)$ is the indicator function of a set $A$ we have that 
$$\prox{\lambda g}{\cdot} = \projop_A(\cdot).$$
This is particularly useful within the \paco framework, as $\projop_\spC(\cdot)$ can be computed efficiently using the stitching trick described in \cref{sec:patch-extraction}, \Cref{eq:proj-op}. However, in general, this trick cannot be applied directly to \cref{eq:paco-gen2}. In order to do that, we need to reformulate~\cref{eq:paco-gen2} so that projecting onto the constraint set, and minimizing the target cost function $f(\cdot)$, can be done separately. One such reformulation is the so called \emph{splitting},
\begin{equation}
(\hvy,\hvz) = \arg\min_{\xi,\zeta} f(\xi) + g(\zeta) + \frac{1}{2\lambda}\|\xi-\zeta\|_2^2\quad\st\quad \xi = \zeta.\label{eq:paco-gen-split}
\end{equation}
The \admm provides a general way to obtain a solution to \cref{eq:paco-gen-split} which is guaranteed to be global if the problem is convex, and local otherwise. The particular \admm steps for this case are described in  \cref{alg:paco}.

\begin{algorithm}
\caption{\label{alg:paco}Generic ADMM program for solving PACO}
\begin{algorithmic}
\REQUIRE $\vz\iter{0} \in \spY$
\REQUIRE $\lambda > 0$
\STATE $t \leftarrow 0$
\STATE $\vu\iter{0} \leftarrow \mathbf{0}$
\REPEAT
\STATE  $\vy\iter{t+1} \leftarrow  \prox{\lambda f}{\vz\iter{t} - \vu\iter{t}}$ \COMMENT{problem dependent:$f$}
\STATE  $\vz\iter{t+1} \leftarrow \Pi_{\spC \cap \Omega}\left(\vy\iter{t+1} + \vu\iter{t}\right)$ \COMMENT{problem dependent:$\Omega$}
\STATE  $\vu\iter{t+1}  \leftarrow  \vu\iter{t} + \vy\iter{t+1} - \vz\iter{t+1}$
\STATE  $t \leftarrow t+1$
\UNTIL{convergence}
\RETURN $\hvy \leftarrow \vy\iter{t}$
\RETURN $\hvz \leftarrow \vz\iter{t}$
\end{algorithmic}
\end{algorithm}

\subsection{Applying PACO}

\paco is a variational method and thus can be fitted to any signal restoration problem that can be written in terms of a cost function and a constraint set. 
This includes the large family of probabilistic methods where $f(\vy_j)$ measures the likelihood of the patch $\vy_j$, either a priori or a posteriori; a few relevant examples of this are~\cite{bayesian-hyperprior,unipriors,paco-dct-ipol,patch-aggregation,gmm,epll}. 

Another group of methods includes penalized likelihood methods where the penalties are not necessarily probabilistic. These include Total Variation~\cite{getreuer} and general linear models based on orthogonal transforms, wavelets~\cite{wavelets1,wavelets2}, learned dictionaries~\cite{ksvd,mod,lewicki99} and sets of linear convolutional filters such as~\cite{convolution1}. Some of these methods consider the energy in terms of the whole image. Still, these images could be seen as patches of a much larger image (for example, in a distributed application for processing huge images, such as those produced by modern scientific telescopes).

In~\cite{manifold1, manifold2} the patches of the image, seen as points in patch space, are assumed to trace a smooth curve over a low-dimensional manifold. Estimations are then obtained by displacing those points so that the resulting curves are smooth.
There is also a large body of works on image processing based on Partial Differential Equations (PDE) which are also variational in nature; see the book~\cite{pde-book} for a complete review of the theory and applications of this approach. More recently, this idea has been combined within the framework of Non-Local methods; the inpainting methods that we use for comparison~\cite{arias,fedorov,newson} all fall into this category.

Finally, other desired features of the patches can be expressed in terms of minimizing a cost function. Examples of this are~\cite{kawai,variational-lce}.\\

As per the constraint sets, these are usually avoided, as dealing with them usually (but not always) leads to slower convergence rates. Perhaps the most common ones are the $\ell_2$ ball constraint, used in some denoising formulations to keep the solutions close to the observed noisy signal, and the inpainting constraint (to be introduced later), which imposes that the results should coincide in those places where the input image is already known. 
The lack of hard constraints besides consensus is good news for \paco, as it tends to complicate the projection stages. As mentioned, fortunately, this is not always the case, as some of our applications show.\\

In order to apply the \paco framework to a particular restoration problem, the user needs to provide two ingredients: the proximal operator of the problem-dependent cost function $f(\cdot)$, and the projection operator onto $\Omega \cap \spC$. If $\Omega = \spY$ then $g=\projop_\spC(\cdot)$ can be efficiently obtained using stitching trick~\cref{eq:proj-op},
$$\vz\iter{t+1} = \oR \left[ \oS \left( \vy\iter{t+1} + \vu\iter{t} \right) \right].$$
If $\Omega \subset \spC$ is affine, the solution can be obtained by iteratively projecting onto $\Omega$ and $\spC$ until convergence (this method is sometimes referred to as POCS). If $\Omega \neq \spC$ is convex but not affine, a general solution can be obtained using Dykstra's algorithm, which we provide in \cref{alg:dykstra} for reference. 
\def\vq{\vec{q}}%
\def\vp{\vec{p}}%
\begin{algorithm}
\caption{\label{alg:dykstra}Dykstra's algorithm for computing the projection of a point $\alpha$ onto the intersection of two sets $A$ and $B$}
\begin{algorithmic}
\REQUIRE point $\alpha$, sets $A$ and $B$
\STATE $t \leftarrow 0$, $\alpha\iter{0} \leftarrow \alpha$
\STATE $\vp\iter{0} \leftarrow \mathbf{0}$, $\vq\iter{0} \leftarrow \mathbf{0}$
\REPEAT
\STATE $\beta\iter{t} \leftarrow \Pi_{B} (\alpha\iter{t}+\vp\iter{t})$
\STATE $\vp\iter{t+1} \leftarrow \alpha\iter{t} + \vp\iter{t} - \beta\iter{t}$
\STATE $\alpha\iter{t+1} \leftarrow \Pi_{A} (\beta\iter{t}+\vq\iter{t})$
\STATE $\vq\iter{t+1} \leftarrow \beta\iter{t} + \vq\iter{t} - \alpha\iter{t+1}$
\STATE $t \leftarrow t+1$
\UNTIL{convergence}
\RETURN $\alpha\iter{t}$
\end{algorithmic}
\end{algorithm}
%
\subsection{Particular case: linear patch models}

\emph{Note that, starting in this section, we well switch to the patch matrix notation defined in \cref{sec:intro}, where patches are arranged as columns of a matrix $\mY$. The same goes to the corresponding coefficient vectors, and all auxiliary variables.}

In many patch-based algorithms, patches are modeled in terms of some linear transformation, i.e., $\mY = \mD\mY$, where $\mD$ is some matrix, and the cost function is a prior defined over the coefficients matrix $\mA$. Typical cases include the \acro{FFT} or \acro{DCT}, and some wavelet transforms. The corresponding splitting formulation is given in terms of the coefficients matrix $\mA$ and its split variable $\mB$: 

\begin{equation}
(\hat{\mA},\hat{\mZ}) = \arg\min_{\mA,\mB} f(\mA) + g(\mD\mB) + \frac{1}{2\lambda}\|\mA-\mB\|_F^2\quad\st\quad \mA = \mB,\label{eq:paco-linear-split}
\end{equation}

Let $g'(\mB) \stackrel{\Delta}{=} g(\mD\mB)$ . If $\mD$ is orthogonal, it is easy to show that $\prox{{\lambda}g'}{\cdot}= \mD\inv \prox{{\lambda} g}{\mD\cdot}.$ This allows us to apply such patch models with only a minor modification to the general solution given in \cref{alg:paco}. The \paco-\dct inpainting problem described in \cref{sec:inpainting} shows an example of exploiting the orthogonality of the \acro{dct} type II.

\def\mP{\mat{P}}

Many recent patch-based algorithms consider the more general situation where $\mD$ is non-orthogonal. A notable case is the use of sparse linear models defined in terms of over-complete dictionaries where $\mD \in \reals^{m{\times}p}$ with $p > m$. Imposing the consensus constraint directly on the dictionary coefficients $\va_j$ as has been proposed in previous works such as~\cite{epll,figueiredo,global-local}, leads to a projection operator which is unwieldy to work with even for very small signals and dictionaries. Referring back to \cref{eq:ops}, we have $\hvx_{ls} = (\mR\transp\mR)\inv \mR\transp \vecop(\mD\hmA)$. Now, under the consensus constraint, we must have $\mR_j\hvx = \mD\hva_j$ for all $j$ or, equivalently, $\mR\hvx = \vecop(\mD\hmA)$. Combined with \cref{eq:ops} we obtain:
\begin{eqnarray*}
\vecop(\mD\hmA) & = & \mR(\mR\transp\mR)\inv \mR\transp \vecop(\mD\hmA)\\
(\mR\transp\mR)\mB\vecop(\hmA) & = & \mR\transp \mB \vecop(\hmA)\\
\vecop(\hmA) &=& [(\mR\mB)\transp\mR\mB]\inv (\mR\mB)\transp \mB \vecop(\hmA)\\
\vecop(\hmA) & = & \mP \vecop(\hmA)
\end{eqnarray*}
where $\mB = (\ident_{n{\times}n} \otimes \mD)$ and $\otimes$ is the kronecker product (see~\cite{matrix-cookbook} for this result), and $\mP=[(\mR\mB)\transp\mR\mB]\inv (\mR\mB)$ is a (non-orthogonal) projection matrix. The consensus set is given in terms of $\mA$ as $$\spC_\mA = \setdef{\mA: (\ident - \mP)\mA = 0 },$$ so it is again a linear subspace of the signal coefficients space. However, by inspecting $\mP$, it is also clear that this is a  matrix whose size $mn{\times}mn$ can be extremely large for practical problems (e.g., $n=10^6$, $m=64$). Even considering the redundancy implied by the Kronecker product, and the sparsity of $\mP$, the number of non-zero elements in $\mP$ is typically very large, making it infeasible to compute $\projop(\mA) = \vecop^{-1}( \mP \vecop(\mA))$ directly. 

The issue with the preceding strategy is to attempt to project the coefficients matrix $\mA$ directly onto the consensus constraint.  The solution presented here is based on the  \emph{Linearized \admm} (\ladmm) or \emph{Uzawa's Method}~\cite{uzawa} for solving \cref{eq:paco-gen}; this method constructs the augmented Lagrangian for the
constraint $\mZ=\mD\mA$ and  then solves a linear approximation of it around  $\mZ$ in each iteration. Global convergence is guaranteed as long as  its parameter $\mu$ satisfies $0 < \mu \leq \lambda/\|\mD\|_2^2$. The \ladmm algorithm is given by,
\begin{eqnarray*}
  \mA\iter{t+1} & \leftarrow & \mathrm{prox}_{\mu f}\left[\mA\iter{t} - (\mu/\lambda)\mD\transp(\mD\mA\iter{t} - \mZ\iter{t}+\mU\iter{t})\right]\\
  \mZ\iter{t+1} & \leftarrow & \mathrm{prox}_{\lambda c}\left[\mD\mA\iter{t+1} + \mU\iter{t}\right] \\
  \mU\iter{t+1} & \leftarrow & \mU\iter{t} + \mD\mA\iter{t+1} - \mZ\iter{t+1}.
\end{eqnarray*}
The corresponding \ladmm algorithm for \paco is given by
\begin{eqnarray*}%
\mA\iter{t+1} & \leftarrow & \mathrm{prox}_{\lambda{f}}\left[\mA\iter{t} - (\mu/\lambda)\mD\transp({\hmY\iter{t}} - \mZ\iter{t} + \mU\iter{t}\right] \\
\hmY\iter{t+1} & \leftarrow & {\mD\mA\iter{t+1}}+\mU\iter{t} \\
\mZ\iter{t+1} & \leftarrow & {\oR[\, \oS (\hmY )\, ]}\\%
\mU\iter{t+1} & \leftarrow & \mU\iter{t} + {\hmY\iter{t+1}} - \mZ\iter{t+1}.
\end{eqnarray*}

\section{Inpainting}
\label{sec:inpainting}

\emph{The method described below has already been treated in detail in~\cite{paco-dct,paco-dct-ipol}. Below we provide a summary of the main results found on those publications, in particular for image inpainting. Please see~\cite{paco-dct} for results on audio and video signals. The source code can be found in the \acro{ipol} paper~\cite{paco-dct-ipol}.
}

The task here is to infer the missing values in an input image $\vx$. We require that the samples in the estimated (inpainted) signal $\hvx$ coincide exactly with those of $\vx$ which are known. Let $O$ denote the subset of indexes where $\vx$ is known. The constraint set  is then defined in \emph{signal space} as 
\begin{equation*}
\Omega'=\{\hvx \in \spX: \hx_i = x_i, i \in O\}.
\end{equation*} 
In order to apply \paco, we need to project onto the constraint set $\Omega$ defined in \emph{patch space} $\spY$. In this particular case, it is easy to show that this can be done by first projecting the patches $\hmY$ onto signal space using the average stitching operator $\oS$ defined in \cref{eq:ops}, then projecting the resulting signal estimate $\hvx$, and finally extracting them again using the operator $\oR$:
\begin{equation*}
\prox{{\lambda}g}{\mZ}=\Pi_{\spC \cap \Omega}(\mZ) = \oR \left\{ \projop_\Gamma\left[ \oS(\mZ)\right] \right\}. 
\end{equation*}

Starting from the standard assumption that the \dct coefficients of image patches follow a heavy tailed distribution, we seek a solution $\hvx$ whose patch coefficients $\va_j=\mD\hvy_j$ are most likely to happen under a Laplacian distribution. Here $\mD$ is the 2D orthogonal \dct type II transform. The cost function is the total Laplacian negative log-likelihood  of the coefficients vectors, where each coefficient is allowed its own Laplacian parameter:

\begin{equation*}
f'(\mA) = \sum_{j=1}^{n}\sum_{i=1}^m \omega_{i}|a_{i,j}|.
\end{equation*}

The proximal operator of $f'(\mA)=\sum_{i,j} \omega_{i}|a_{i,j}|$ is known as the \emph{soft-thresholding} operator and is given by \cref{eq:soft-thres},
\begin{equation}
\ost_{\lambda{w}_{i}}(a) = 
\left\{
\begin{array}{ccl}
a + \lambda{w}_{i}&,& a < -\lambda{w}_{i} \\
0&,& -\lambda{w}_{i} \leq a \leq \lambda{w}_{i} \\
a - \lambda{w}_{i}&,& a > \lambda{w}_{i} \\
\end{array}
\right.
\label{eq:soft-thres}
\end{equation}

The \admm solution to the \paco-\dct inpainting problem defined in terms of the cost function and constraint sets is given in \cref{alg:paco-dct-inpainting}.

\begin{algorithm}
\caption{\label{alg:paco-dct-inpainting}Complete \acro{PACO-DCT}algorithm.}
\begin{algorithmic}
\REQUIRE input signal $\vx$, initial estimation $\hvx\iter{0}$, observed samples index set $O$
\STATE  $\mZ\iter{0} \gets \oR(\hvx\iter{0})$; $\mU\iter{0} \gets \mathbf{0}$
\REPEAT
\STATE $\mA\iter{t} \gets \mD\transp[\mZ\iter{t}+\mU\iter{t}]$ \quad\COMMENT{DCT coefficients}
\STATE    $a_{i,j}\iter{t+1}  \leftarrow  \ost_{\lambda \omega_{i,j}} (b_{i,j}\iter{t}  - u_{i,j}\iter{t}),\;\forall\,i,j$ \quad\COMMENT{soft thresholding}
\STATE$\mY\iter{t+1} \gets \mD\mA\iter{t+1}$ \quad\COMMENT{convert to patches}
\STATE$\hvx\iter{t+1}  \leftarrow  \oS[\,\mA\iter{t+1}\,] \nonumber$ \quad\COMMENT{stitch}
\STATE $\hat{x}_i\iter{t+1} \leftarrow x_i\iter{t+1},\;\,\forall\,i\in O$ \quad\COMMENT{Project onto $\Omega'$}
\STATE $\mZ\iter{t+1}  \leftarrow \oR( \hvx\iter{t+1} )$ \quad\COMMENT{extract patches}
\STATE $\mU\iter{t+1}  \leftarrow  \mU\iter{t} + \mY\iter{t+1} - \mZ\iter{t+1}$ 
\STATE $t \gets t+1$
\UNTIL{convergence}
\RETURN inpainted signal $\hvx\iter{t}$
\end{algorithmic}
\end{algorithm}

\subsection{Sample results}
\label{sec:inpainting-results}

The results below are taken from~\cite{paco-dct-ipol}. The full results can be found in \url{http://iie.fing.edu.uy/~nacho/paco/results}.
\begin{figure*}
\includegraphics[width=0.24\textwidth]{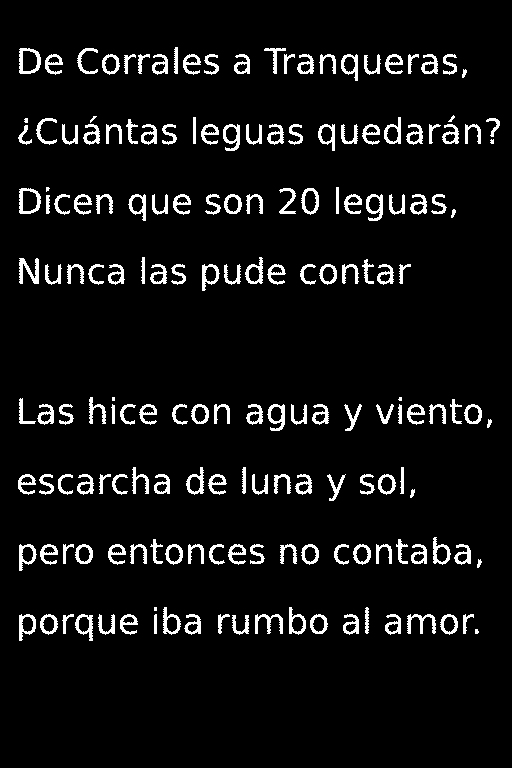} %
\includegraphics[width=0.24\textwidth]{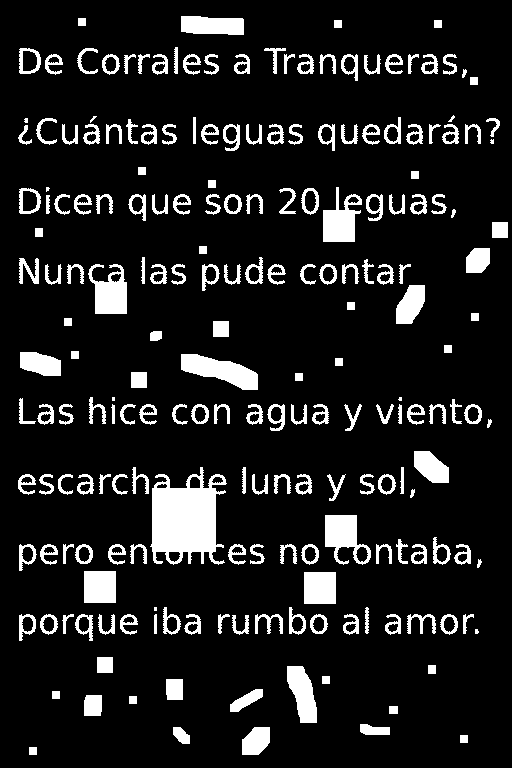} %
\includegraphics[width=0.24\textwidth]{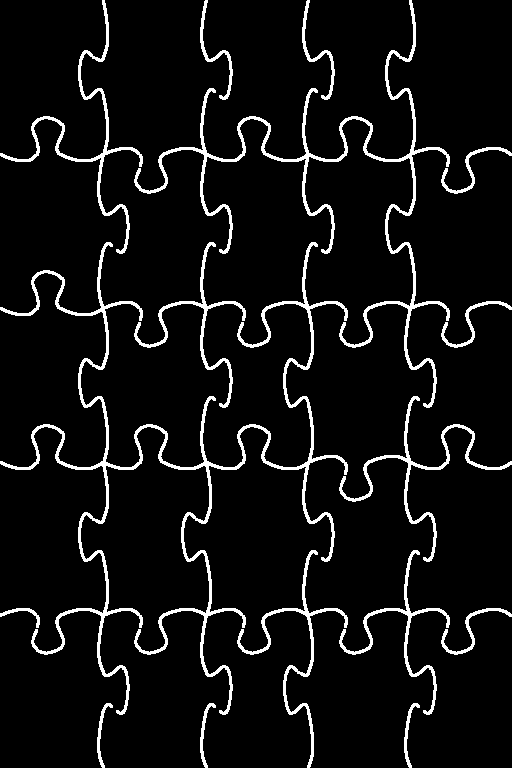} %
\includegraphics[width=0.24\textwidth]{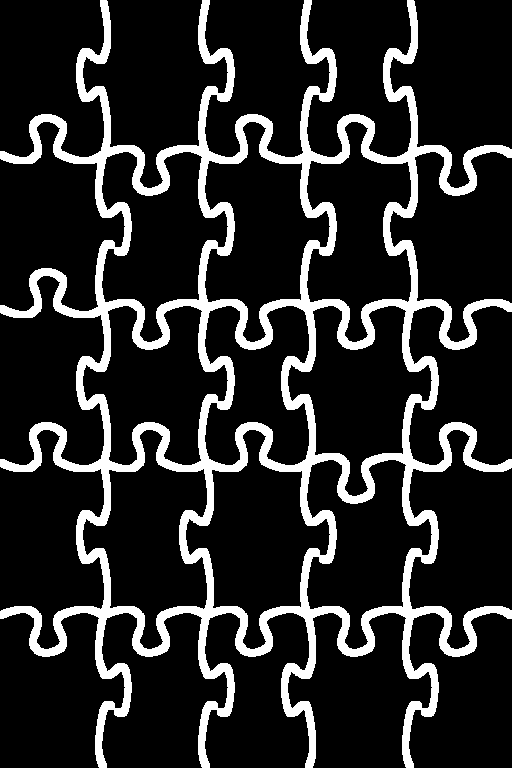}\\[0.5ex]%
\includegraphics[width=0.24\textwidth]{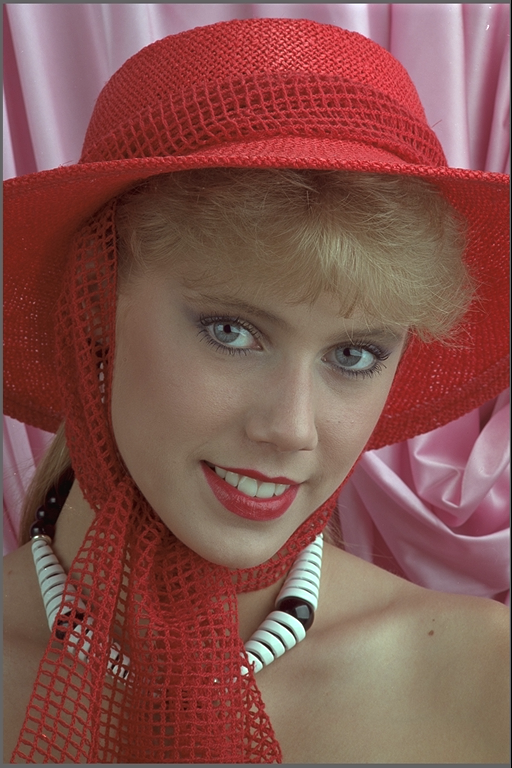} %
\includegraphics[width=0.24\textwidth]{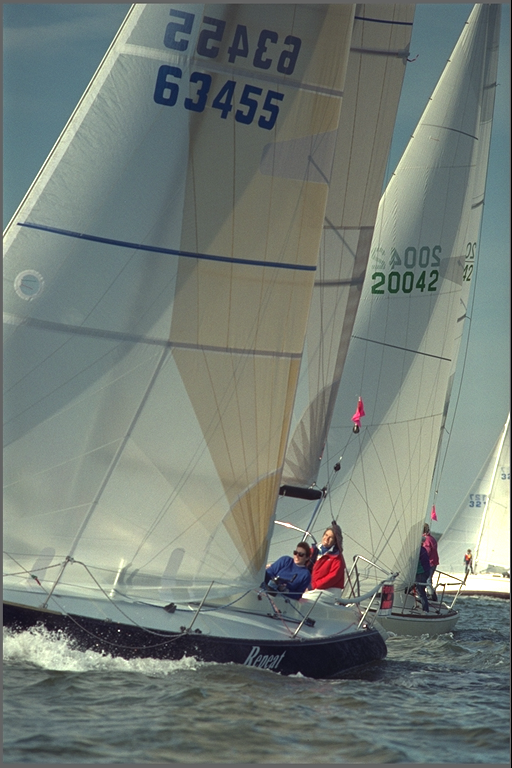} %
\includegraphics[width=0.24\textwidth]{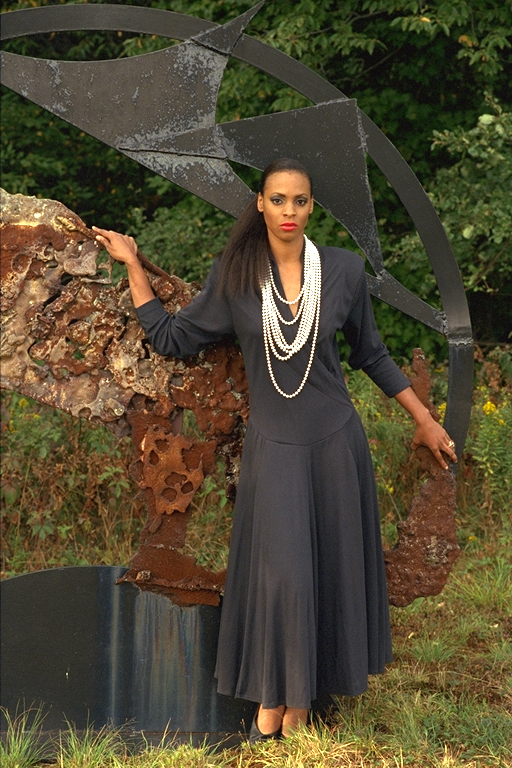} %
\includegraphics[width=0.24\textwidth]{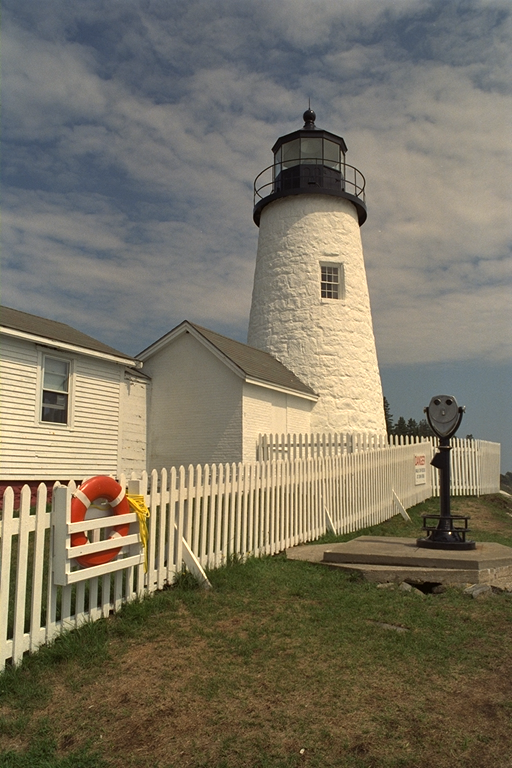}
\caption{\label{fig:masks}Inpainting masks 1 to 4. Masks 2 and 4 are more challenging due to the size of the erasures.}
\end{figure*}

\Cref{fig:masks} shows the different missing pixels masks, and sample images from the Kodak dataset on whose 24 images we tested our algorithms using those four masks. Our results are compared to~\cite{maerz,fedorov,newson}. 

We use the \paco-\dct problem parameters that yield the best compromise between \acro{rmse} and \acro{ssim}: $w=16$ ($m=256$) and $s=2$. The optimization parameters are set to $\lambda=10$, $\kappa=0.95$ a maximum of $256$ iterations and a minimum allowable cost function decrease of $\epsilon=10^{-5}$. 
\renewcommand{\arraystretch}{1.2}
\def\B{\bf\color{Purple}}
\begin{table}\caption{\label{tab:kodak} Summary of inpainting results on the whole Kodak dataset (luminance channel) in terms of \acro{rmse} (smaller is better) and \acro{ssim} (higher is better); best results are in bold purple.} 
\tabcolsep 6pt%
\vspace{2ex}
\centering\begin{tabular}{|c|cccc|cccc|}\hline
  \acro{METRIC}$\rightarrow$	& \multicolumn{ 4 }{|c|}{ \acro{rmse} } & \multicolumn{ 4 }{|c|}{ \acro{ssim} } 	\\ \hline
  \acro{MASK}$\rightarrow$	& 1 & 2 & 4a & 4c & 1 & 2 & 4a & 4c 	\\ \hline
  \acro{METHOD}$\downarrow$	& \multicolumn{ 8 }{|c|}{\acro{PERCENTILE} 25 } 	\\ \hline
\paco	        & 	\B8.38	&	15.68	&	\B8.15	&	\B11.56	&	\B0.9886	&	\B0.9562	&	\B0.9879	&	\B0.9513	\\
\cite{fedorov}	&   10.44	&	\B15.07	&	9.80	&	12.92	&	0.9868	&	0.9556	&	0.9863	&	0.9392	\\
\cite{newson}	&	12.12	&	19.64	&	10.88	&	14.77	&	0.9825	&	0.9418	&	0.9826	&	0.9299	\\
\cite{maerz}	&	19.84	&	23.67	&	20.68	&	23.90	&	0.8791	&	0.8548	&	0.8628	&	0.8403	\\
  \hline
  METHOD	& \multicolumn{ 8 }{|c|}{\acro{PERCENTILE} 50 } 	\\ \hline
\paco	&	\B10.48	&	17.32	&	\B10.05	&	\B14.09	&	\B0.9906	&	\B0.9615	&	\B0.9900	&	\B0.9567	\\
\cite{fedorov}	&	12.46	&	\B16.31	&	12.08	&	15.94	&	0.9886	&	0.9607	&	0.9888	&	0.9508	\\
\cite{newson}	&	14.75	&	22.48	&	13.91	&	17.96	&	0.9856	&	0.9463	&	0.9846	&	0.9389	\\
\cite{maerz}	&	24.15	&	29.25	&	26.21	&	29.23	&	0.9103	&	0.8854	&	0.9084	&	0.8777	\\
  \hline
  METHOD	& \multicolumn{ 8 }{|c|}{\acro{PERCENTILE} 75 } 	\\ \hline
\paco	&	\B13.43	&	\B20.73	&	\B15.20	&	\B19.66	&	\B0.9931	&	0.9646	&	\B0.9918	&	\B0.9618	\\
\cite{fedorov}	&	15.95	&	21.20	&	17.65	&	20.90	&	0.9914	&	\B0.9660	&	0.9910	&	0.9571	\\
\cite{newson}	&	18.92	&	27.20	&	20.87	&	24.72	&	0.9887	&	0.9494	&	0.9886	&	0.9499	\\
\cite{maerz}	&	31.05	&	37.91	&	32.12	&	36.17	&	0.9414	&	0.9131	&	0.9378	&	0.9087	\\
  \hline
  \end{tabular}
\end{table}

\begin{figure}
\centering\includegraphics[height=0.24\textheight]{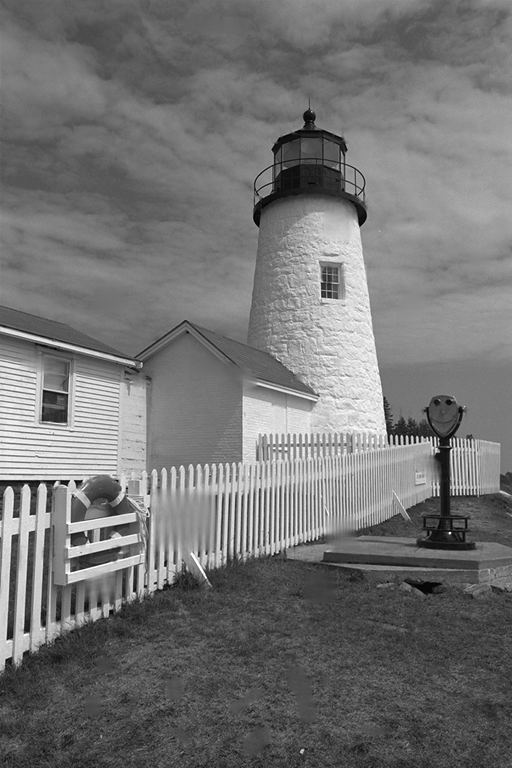}\hspace{0.25ex}%
\centering\includegraphics[height=0.24\textheight]{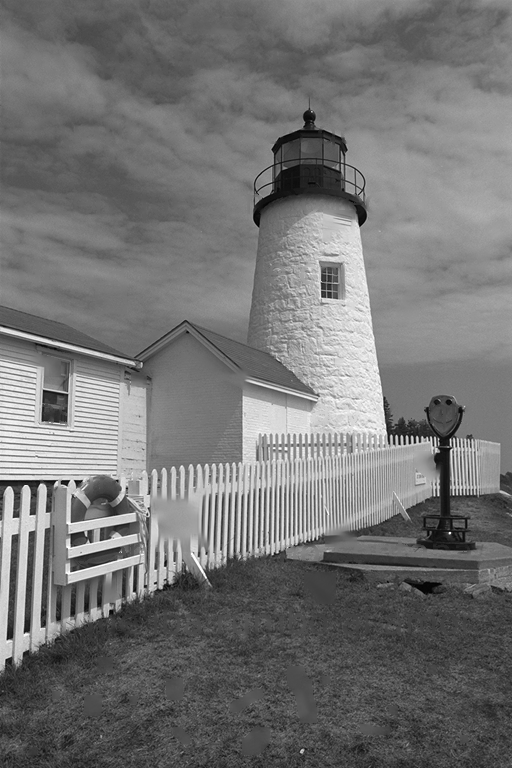}%
\centering\includegraphics[height=0.24\textheight]{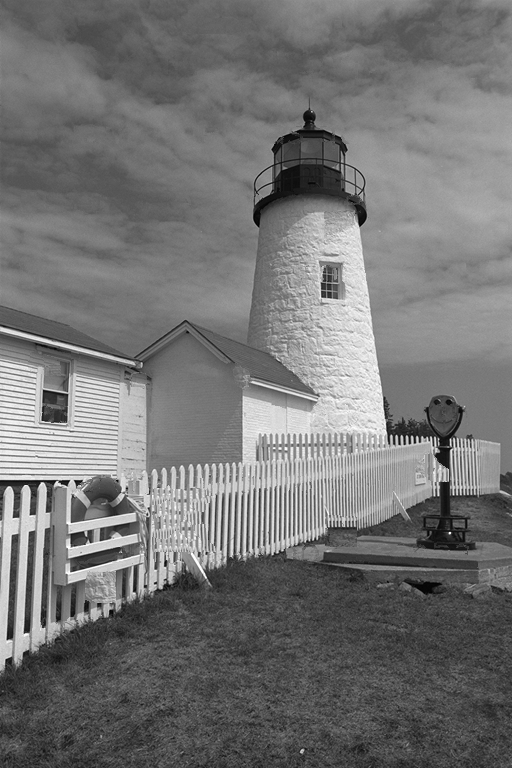}\hspace{0.25ex}%
\centering\includegraphics[height=0.24\textheight]{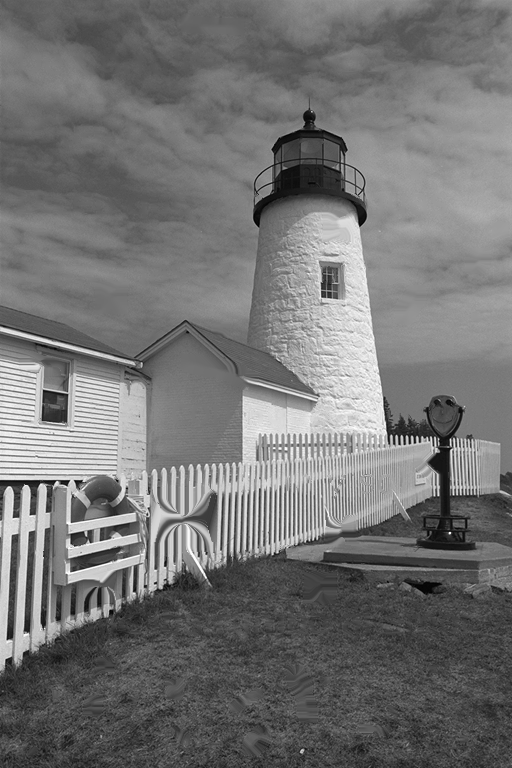}%
\caption{%
\label{fig:visual-results}
Visual comparison of the four methods on Kodak \#19 and mask \#2; please enlarge page to view details. Top to bottom, left to right: \paco, \cite{maerz,fedorov,newson}. All four methods perform well on narrow gaps. For wider gaps, 
the results of \cite{fedorov} and \acro{paco-dct} are similar, the major difference being the fence next to the life saver; \cite{fedorov} produces sharper results than \paco, but also more artifacts (e.g., large gap in fence or above the lighthouse window); \cite{newson} and \cite{maerz} produce significant artifacts in these regions (we recall the reader that \cite{maerz} is not suited to such large gaps by design.)
}%
\end{figure}
\Cref{tab:kodak} summarizes the results of the four methods for the 24 Kodak images in terms of the median, 25 and 75 percentiles of the \acro{rmse} and \acro{ssim} measures, again computed \emph{only} over the missing pixel locations. For simplicity, we  only show the results on the  Luminance channel of these images; similar results are obtained for the other channels. \Cref{fig:visual-results} shows the results obtained with these four methods for a particular case, again on the Luminance channel.

\section{Denoising}
\label{sec:denoising}
\def\Gaussian{\mathcal{N}}
\def\gmm{\textsc{gmm}\xspace}
\def\map{\textsc{map}\xspace}
\def\sure{\textsc{sure}\xspace}
\def\epll{\textsc{epll}\xspace}
\def\rmse{\textsc{rmse}\xspace}
\def\dj{\textsc{dj}\xspace}
In this section we apply the \paco framework to the classic problem of removing i.i.d. Gaussian noise of known variance from images. 
The goals here are to demonstrate: a) the simplicity of applying \paco to already existing formulations and b) that doing so has the potential of improving their results. Please visit~\url{http://iie.fing.edu.uy/~nacho/paco/} for source code and the full set of results.

We now provide a brief introduction to the problem, and a few key references to works directly related to our approach. The input is a noisy signal $\vx=\tvx + \eta$, where $\eta \sim \Gaussian(0,\sigma^2)$, and the task is to estimate the clean, unobserved signal $\tvx$ from $\vx$.
Typical approaches involve some combination of a priori information characteristic of clean signals, such as a prior distribution $p(\vx)$, and the fact that the noise is i.i.d. with mean zero, so that it can be canceled by adding noisy samples together. Prior-based approaches lend naturally to a variational formulation by defining a cost function $f(\vx) = -\log p(\vx)$; they usually differ on how they take into account the noise process. For instance, one might impose that the recovered signal is no farther away from $\vx$  than its expected $\ell_2$ distance from the clean one:
\begin{equation*}
\hvx = \arg\min_\zeta f(\zeta)\;\st\;\|\vx-\zeta\|_2^2 \leq \sigma^2.
\end{equation*}
Another popular approach is the classic Maximum a Posteriori (\map) estimator, which leads to an unconstrained problem. This is the case of works such as~\cite{elad06}:
\begin{equation*}
\hvx = \arg\min_\zeta f(\zeta) + \frac{1}{2\sigma^2}\|\vx-\zeta\|_2^2.
\end{equation*}

Besides \map, other notable successful estimators are the Stein's Unbiased Estimator (\sure)~\cite{sure}, and Donoho \& Johnstone's shrinkage operator (\dj)~\cite{donoho94}. Many patch-based denoising methods translate these same methods to patch space by considering each patch to be a small noisy image, for example,  Donoho-Johnstone's shrinkage is applied in~\cite{patch-dct-ipol} to \dct patch coefficients, \sure is employed in~\cite{patch-sure}, and the \map approach is used in \cite{elad06,epll}. For instance, the patch-based \map denoising formulation can be written as:
\begin{equation*}
\hvy = \arg\min_\zeta \sum_{j=1}^{n}\left\{ f(\zeta_j) + \frac{1}{2\sigma^2}\|\vy_j-\zeta_j\|_2^2\right\},
\end{equation*}
where $\zeta_j$ are the individual patches and $\zeta$ is the whole patches vector over which we minimize the cost.

Since we want to show how \paco can result in \emph{relative} improvements over already-existing patch-based denoising methods, our ideal starting point would be a reasonably recent and performant \map-based method. Ideally, too, the prior should be convex, so that results do not depend on, for example, different initialization strategies. Two good candidates are \gmm~\cite{gmm} and \epll~\cite{epll}: although neither are convex, both are recent, competitive, and fall within the patch-based \map denoising framework. Of both, the \epll has the advantage of using a pre-learned prior which we can use as well, thus allowing us to concentrate more on the  effects of the consensus constraint. Moreover, \epll also includes a for of \emph{soft consensus} between the patches, which also serves us to compare the effect of different consensus strategies within the same setting.

We now provide a brief introduction to \epll, and then develop a variant of the algorithm using \paco instead of the soft consensus just mentioned.

\subsection{Expected patch log-likelihood}

The \epll framework was introduced in~\cite{epll}. Formally, the \epll acronym refers to the common, aforementioned idea of imposing a probabilistic prior on patches and then producing an estimate of the restored signal whose patches are the most likely to occur given the observed, degraded signal. The \epll algorithm is a variational, patch-based method which uses a pre-learned Gaussian Mixture Model (\gmm) such as the one introduced in~\cite{gmm} and assumes an i.i.d. Gaussian model of known variance for the image noise, that is:
\[
\vx = \tvx + \eta,\quad\eta\sim \Gaussian(0,\sigma^2),\;\tvx_j\sim p(\tvx) 
\]
where
\[
p(\tvx_j) = w_{k_j} \mathcal{N}(0,\Sigma_k),\;k_j=1,\ldots,K,
\]
where $K$ is the number of modes in the \gmm model.

The denoising procedure is essentially a straightforward \map estimation of the patches followed by standard patch averaging.
The key contribution of~\cite{epll} is the realization that, after patch averaging, the resulting patches in the denoised image are no longer, individually, the most likely ones given the prior (for example, they may be overly smooth). Therefore, the authors include a quadratic term in the formulation which encourages the overlapped patches to be likely too. The target energy to be minimized is expressed as a function of the denoised image estimate $\hvx$ as:
\begin{equation}
\sum_{j=1}^{n} f(\mR_j\hvx) + \frac{1}{2\sigma^2}\sum_{i=1}^{N}o_{i}\|\hvx_i-\vx_i\|_2^2,
\label{eq:epll-0}
\end{equation}
where $f(\mR_j\hvx)=-\log p(\mR_j\hvx)$ is the log-prior of the $j$-th patch of the image and $o_i$ is the number of patches that the $i$-th sample of $\vx$ is mapped to. Following our notation conventions, this corresponds to the $i$-th element of the diagonal of $\mR\transp\mR$, $o_i=(\mR\transp\mR)_{ii}$. 

\subsection{The EPLL algorithm}

Directly minimizing \cref{eq:epll-0} on $\hvx$ runs into the same difficulties that were mentioned \cref{sec:patch-methods}, namely, they involve a huge projection matrix. The authors of \epll~\cite{epll} sidestep this problem using a splitting method proposed in~\cite{half-quadratic-splitting}, which can be seen as a \emph{soft} version of our proposed \emph{hard} splitting \cref{eq:paco-gen-split}. Auxiliary (split) variables $\{\hvz_1,\hvz_2,\ldots\hvz_n\}$ are defined, each corresponding to a patch on the image, and a new (not equivalent) problem is solved instead:
\begin{equation}
\sum_{j=1}^{n} f(\hvz_j) + \frac{1}{2\sigma^2}\sum_{i=1}^{N}o_{i}\|\hvx_i-\vx_i\|_2^2 + \frac{\beta}{2\sigma^2}\sum_{j=1}^{n}\|\mR_j\hvx-\hvz_j\|_2^2,
\label{eq:epll-split}
\end{equation}
where $\beta$ is a penalty term. The energy \cref{eq:epll-split} is then minimized alternatively on the split $\{\hvz_1,\hvz_2,\ldots\hvz_n\}$ and main ($\hvx$) variables for a fixed number of iterations, each with an increasing value of the penalty $\beta$; based on experimentation, the authors of~\cite{epll}  propose six iterations with $\beta$ taking on the values $(1,4,8,16,32,64)$. As $\beta$ is increased, the patches of the image iterate $\hvx$ and their split variables are encouraged to match, $\mR_j\hvx \approx \hvz_j$; this is a form of soft consensus which converges to exact consensus for $\beta\rightarrow\infty$.

We now reformulate the problem in a form which is more consistent with our notations and conventions by writing \cref{eq:epll-split} in terms of the patches vectors of the input (noisy) image, $\vy=\mR\vx$, and its denoised estimate, $\hvy=\mR\hvx$, and $\tau=\sigma^2/\beta$:
\begin{equation}
\sum_{j=1}^{n} \left[ f(\hvz_j) + \frac{1}{2\sigma^2}\|\hvy_j-\vy_j\|_2^2 + \frac{1}{2\tau}\|\hvy_j-\hvz_j\|_2^2 \right].
\label{eq:epll-split-1}
\end{equation}
Also, for convenience, we define the patches vector $\hvz=[\hvz_1|\hvz_2|\ldots| \hvz_n ]$. The alternate minimization proceeds as follows. For fixed $\hvz$, \cref{eq:epll-split-1} is minimized with respect to $\hvy$. This is a simple quadratic problem whose solution is a weighted sum of the noisy input image $\vx$ and the image obtained by average stitching the auxiliary patches $\hvz$:
\begin{equation}
\hvy\iter{t+1} = \arg\min_\zeta  
\frac{1}{2\sigma^2}\|\zeta-\mR\vx\|_2^2 +
\frac{1}{2\tau}\|\zeta-\hvz\iter{t}\|_2^2  = \frac{\vx+\beta\oS(\hvz\iter{t})}{1+\beta}.
\label{eq:epll-stitch}
\end{equation}
For fixed $\hvy\iter{t+1}$, the solution to~\cref{eq:epll-split-1} with respect to $\hvz$ corresponds to the proximal operator of $f(\cdot)$ given the parameter $\tau=\beta/\sigma^2$, which is separable in the patches:
\begin{equation}
\hvz_j\iter{t+1} = \arg\min_\xi
f(\xi) + \frac{1}{2\tau}\|\xi-\hvy_j\iter{t+1}\|_2^2,\;\forall\,j=1,\ldots,n.
\label{eq:epll-prox-0}
\end{equation}

The \gmm prior is non-convex and so \cref{eq:epll-prox-0} can only be solved approximately. The \epll algorithm does so by first assigning each patch $\hvy_j\iter{t+1}$ to the Gaussian mode $k_j$ with maximum posterior probability given $\hvy_j\iter{t+1}$:
\begin{equation}
k_j = \arg\max_k \log p(k|\hvy_j\iter{t+1}) 
= \arg\max_k 
\left\{ 
\log w_k - \frac{1}{2}\log|(\Sigma_k+\tau\mI)| - 
\frac{1}{2}\zeta\transp(\Sigma_k+\tau\mI)\inv\zeta
\right\},
\label{eq:epll-approx-map}
\end{equation}
where $\Sigma_k$ is the covariance matrix of the $k$-th Gaussian mode of the \gmm (the mean of all the modes is assumed to be $0$).
Given $k_j$, the problem becomes quadratic and the patch estimate can be obtained in closed form:
\begin{equation}
\hvz_j\iter{t+1} = \arg\min_\xi
\frac{1}{2}\xi\transp\Sigma_{k_j}\inv\xi + \frac{1}{2\tau}\|\xi-\hvy_j\iter{t+1}\|_2^2 = (\Sigma_{k_j} + \tau\mI)\inv\Sigma_{k_j} \hvy_j\iter{t+1}.
\label{eq:epll-prox}
\end{equation}

The \epll  method is summarized in \cref{alg:epll}.
\begin{algorithm}
\caption{\label{alg:epll}Alternate block minimization -- EPLL algorithm}
\begin{algorithmic}
\REQUIRE noisy image $\vx$, noise variance $\sigma^2$
\STATE $t \leftarrow 0$ \;
\STATE $\hvx\iter{0} \leftarrow \vx$
\FOR {$\beta=1,4,8,16,32,64$}
\STATE  $\tau\iter{t} = \sigma^2/\beta$
\STATE  $\hvy\iter{t} \leftarrow \mR\hvx\iter{t}$
\STATE  $k_j = \arg\min_j -\log w_k + \frac{1}{2}\log|\Sigma_k+\tau\iter{t}\mI| + \frac{1}{2}(\hvy_j\iter{t})\transp\left[\Sigma_k+\tau\iter{t}\mI\right]\inv\hvy_j\iter{t}$
\STATE  $\hvz_j\iter{t+1} \leftarrow (\Sigma_{k_j} + \tau\iter{t}\mI)\inv\Sigma_{k_j} \hvy_j\iter{t}$
\STATE  $\hvx\iter{t+1} \leftarrow (1+\beta)^{-1}\left[\vx+\beta(\mR\transp\mR)\inv\mR\transp\hvz\iter{t+1}\right]$
\STATE $t \leftarrow t+1$
\ENDFOR
\RETURN denoised image $\hvx\iter{t}$
\end{algorithmic}
\end{algorithm}

\subsection{Learned vs fixed GMM model}

The above EPLL method is defined for a pre-learned GMM model. However, both in~\cite{epll,epll-ipol} the implementations allow for the input model to be further adapted using the standard Expectation-Maximization algorithm for GMM models~\cite{gmm}. The denoising method that we present below, as well as the EPLL results that we obtained using the implementation~\cite{epll-ipol}, do not adapt the GMM model. This allows us to perform a comparison of both methods where the only differences lie in the splitting methods used (\cite{half-quadratic-splitting} vs \admm).
This being said, as with \epll, nothing prevents our algorithm to update the \gmm model during the iterations.

\subsection{PACO-GMM}

By now it should be clear that, for this particular case, \epll bears many similarities with \paco: it is a variational patch-based method which, albeit indirectly, encourages consensus between the estimated patches. Last but not least, the numerical solution is based on a splitting strategy. 
We now develop the ``\paco equivalent'' to the \epll method, that is, we employ the same prior model and the same approximate method for estimating a denoised patch given a noisy observation~\cref{eq:epll-approx-map} and~\cref{eq:epll-prox}.
Concretely, the \paco formulation for the \map-denoising problem is given by:
\begin{equation}
\hvy = \arg\min_{\zeta} f(\zeta) + \frac{1}{2\sigma^2}\|\zeta-\vy\|_2^2 + g(\zeta),
\label{eq:paco-denoising}
\end{equation}
where $g(\zeta)$ is the consensus constraint and $\vy$ are the patches extracted from the input noisy image. It is interesting to note that the splitting in this case can be done in two ways: the quadratic term in~\cref{eq:paco-denoising} can be merged with either $f(\cdot)$ or $g(\cdot)$. In the first case we obtain:
\begin{equation}
(\hvy,\hvz) = 
\arg\min_{(\zeta,\xi)} 
\left[ 
f(\xi) 
+ \frac{1}{2\sigma^2}\|\zeta-\xi\|_2^2
\right]
+ g(\zeta) 
+ \frac{1}{2\tau}\|\xi-\vy\|_2^2,\quad\st\;\zeta=\xi;
\label{eq:paco-denoising-1}
\end{equation}
and in the second case
\begin{equation}
(\hvy,\hvz) = \arg\min_{(\zeta,\xi)} f(\xi) + 
\left[
g(\zeta) 
+ \frac{1}{2\sigma^2}\|\zeta-\vy\|_2^2
\right]
+ \frac{1}{2\tau}\|\zeta-\xi\|_2^2,\quad\st\;\zeta=\xi.
\label{eq:paco-denoising-2}
\end{equation}
Despite being formally equivalent, \cref{eq:paco-denoising-1} and \cref{eq:paco-denoising-2} admit different interpretations. 
In \cref{eq:paco-denoising-1}, the first cost function is the log-posterior of the patch estimate given the corresponding noisy patch, and the second function is the consensus indicator. In \cref{eq:paco-denoising-2}, the cost function is the log-prior of the patch estimate, and the quadratic fitting term is absorbed into the consensus indicator function. 

We now observe that, if we disregard the equality constraint, the solution of \cref{eq:paco-denoising-2} for fixed $\xi$ with respect to $\zeta$ can be manipulated to yield a familiar form:
\begin{eqnarray}
\hvy &=& \arg\min_{\zeta} g(\zeta) + \frac{1}{2\sigma^2}\|\zeta-\vy\|_2^2 + \frac{1}{2\tau}\|\zeta-\xi\|_2^2 \nonumber\\
 &=& \arg\min_{\zeta} g(\zeta) + \frac{\tau+\sigma^2}{2\tau\sigma^2}\left\|\zeta-\frac{\tau\vy+\sigma^2\xi}{\tau+\sigma^2}\right\|_2^2\nonumber\\
&=& \arg\min_{\zeta} g(\zeta) + \frac{1+\beta}{2\sigma^2}\left\|\zeta-\frac{\vy+\beta\xi}{1+\beta}\right\|_2^2\nonumber\\
&=& \oR\left[\oS\left(\frac{\vy+\beta\xi}{1+\beta}\right)\right] = \hvy_{ls}.
\label{eq:paco-denoising-stitch}
\end{eqnarray}
The resulting expression~\cref{eq:paco-denoising-stitch} is actually equivalent to~\cref{eq:epll-stitch}. Furthermore, if we fix $\hvy$ in \cref{eq:paco-denoising-2} and solve for $\hvz$ we arrive exactly at \cref{eq:epll-prox}! 

Based on the above facts, we choose to follow our algorithm along the splitting defined in~\cref{eq:paco-denoising-2}. Not only this leads to a solution that is as close as possible to that of \epll, but allows us to use the approximate \map estimation proposed in~\cite{epll}, which is the most delicate part of the process, out of the box. Of course, our splitting does not operate directly on $\hvx$ and $\hvy$ but includes the Lagrangean multiplier; this amounts to solving \cref{eq:paco-denoising-stitch} with $\vy+\vu$ in place of $\vy$, \cref{eq:epll-prox} with $\vz-\vu$ instead of $\vz$, and then updating $\vu \leftarrow \vu + \vy - \vz$. 

There is another important aspect in this case which is the role of the \admm penalty $\tau$. In standard \admm this value is kept fixed, but there is no harm in letting it grow with the iterations. However, care must be taken if (as we did so far) the Lagrangian multiplier $\vu$ is pre-scaled by $\tau$: the algorithm breaks down if $\tau$ varies during iterations. 
Now, for a highly non-convex problem like \cref{eq:paco-denoising}, the choice of $\tau$, and whether to increase it or not during iterations, may be crucial for converging to a good local minima. 
A deep analysis of the numerical implications of this choice is clearly beyond this work. Instead, again, we resort to the heuristics used in \epll in the following way: the \epll algorithm is run for $7$ fixed iterations, for $\beta=1,2,4,8,16,32,64$; these values were found empirically to work best for most noise levels. As we saw in our analysis of the \epll Algorithm, the nexus between the \admm penalty parameter and $\beta$ is given by $\tau=\beta/\sigma^2$. Thus, we follow a similar strategy and apply our algorithm for varying values of $\tau$. In our case, we found that $\tau=i^2,i=1,2,3,4,5,6,\ldots$, where $i$ is the iteration number, gives excellent results on par or better than \epll, as we will see below. 

The resulting \paco-\gmm is given in Algorithm~\cref{alg:paco-gmm}. We show the differences between both methods in purple for better comparison.

\begin{algorithm}
\caption{\label{alg:paco-gmm}PACO-GMM algorithm. Differences with the EPLL \cref{alg:epll} are marked in purple.}
\begin{algorithmic}
\REQUIRE noisy image $\vx$, noise variance $\sigma^2$, {\color{Purple} \admm parameter $\tau$}
\STATE $t \leftarrow 1$
\STATE $\hvy\iter{t} \leftarrow \mR\vx$
\STATE $\vu\iter{t} \leftarrow \mathbf{0}$
\WHILE{{\color{Purple}$t \leq 8$}}
\STATE  $\beta=t^2$
\STATE  $\tau  \leftarrow \sigma^2/\beta$ 
\STATE  $k_j = \arg\min_j -\log w_k + \frac{1}{2}\log|\Sigma_k+\tau\iter{t}\mI| + \frac{1}{2}(\hvy_j\iter{t}{\color{Purple}-\tau\vu\iter{t}})\transp\left[\Sigma_k+\tau\iter{t}\mI\right]\inv(\hvy_j\iter{t}{\color{Purple}-\tau\vu\iter{t}})$
\STATE  $\hvz_j\iter{t+1} \leftarrow (\Sigma_{k_j} + \tau\iter{t}\mI)\inv\Sigma_{k_j} (\hvy_j\iter{t}{\color{Purple}-\tau\vu\iter{t}})$   
\STATE  $\hvx\iter{t+1} \leftarrow (1+\beta)^{-1}\left[\vx+\beta(\mR\transp\mR)\inv\mR\transp(\hvz\iter{t+1}{\color{Purple}+\tau\vu\iter{t}})\right]$ 
\STATE  $\color{Purple}\vu\iter{t+1} \leftarrow \vu\iter{t} + (1/\tau)(\mR\hvx\iter{t+1} - \hvz\iter{t+1})$
\STATE  $t \leftarrow t+1$
\ENDWHILE
\RETURN denoised image $\hvx$
\end{algorithmic}
\end{algorithm}

\subsection{Qualitative Comparison}

As expected and, by comparing \cref{alg:epll} and \cref{alg:paco-gmm}, it is clear that both methods are very similar. The main difference, of course, lies in the use of a hard constraint instead of a soft one, which in turns leads to the presence of the Lagrangian multiplier $\vu$ in \cref{alg:paco-gmm}.
Moreover, note that, as the Lagrange multiplier is initialized to $\vu\iter{1}=0$, the first iteration of both methods is \emph{exactly} the same. However, the iterates in both methods diverge  after the first iteration, precisely because of $\vu\iter{t} > 0$ for $t>1$. Despite this divergence, both methods yield very similar results in practice, particularly from the quantitative point of view, as shown in \cref{tab:denoising-results}.

\subsection{Variants using other estimators}

Both \epll and our \paco-\gmm method estimate the denoised patches using the \map corresponding to a Gaussian prior on the patches; this corresponds to an $\ell_2$-penalized least squares problem. However, nothing prevents us to apply a different estimator (keeping the mode assignment intact), as done for example in~\cite{gmm}. Concretely, we replace the $\ell_2$ estimator \cref{eq:epll-prox}, first with an $\ell_1$ prior, and second  with the \dj hard thresholding estimator.  In the former case we have,
\def\vs{\mathbf{s}}
\def\mV{\mathbf{V}}
\begin{equation}
\hvz_j\iter{t+1} = \arg\min_\xi
\sum_i \frac{1}{s_i}|(\mV\transp\xi)_i| + \frac{1}{2\tau}\|\xi-\hvy_j\iter{t+1}\|_2^2 
= \mV\mathcal{T}_{\tau/\vs}(\mV\transp\hvy_j\iter{t+1}),
\label{eq:l1-denoiser}
\end{equation}
where $\mV\mathrm{diag}(\vs)\mV\transp = \Sigma_k$ is the Singular Value Decomposition of the covariance matrix, $\vs$ is the vector of singular values of that decomposition, and $\mathcal{T}_{\tau/vs}(\cdot)$ is the soft-thresholding operator with threshold $\tau/s_i$ applied to the component $i$ of the input vector, which we repeat here for convenience:

$$\mathcal{T}_{\tau/\vs}(\xi)_i = 
\left\{
\begin{array}{ccl}
-\xi + \tau/s_i &, & \xi < -\tau/s_i\\
0 &, & -\tau/s_i \leq \xi \leq \tau/s_i\\
 \xi - \tau/s_i &, & \xi > \tau/s_i\\
\end{array}
\right..
$$

As for the latter, the estimate is obtained as $\mV\mathcal{H}_{3\sigma}(\mV\transp\hvy_j\iter{t+1})$ with:
\begin{equation}
\mathcal{H}_{3\sigma}(\xi)_i = 
\left\{
\begin{array}{ccc}
\xi &, & |\xi| >  3\sigma\\
0 &, & |\xi| \leq 3\sigma \\
\end{array}
\right..
\label{eq:dj-denoiser}
\end{equation}
where the threshold $3\sigma$ is suggested in~\cite{patch-dct-ipol}.

\def\best{\color{Purple}}
\begin{table}
\caption{\label{tab:denoising-results}Summary of denoising results on the grayscale Kodak images. For each noise level and method we show the 25th, 50th (the median) and 75th percentiles of the \rmse (lower is better). The difference between the medians of all methods in all settings is very small (less than 1 intensity level). In particular, those of \paco-\gmm, its $\ell_1$ variant, and \epll are almost identical. BM3D yields slightly but consistently smaller \rmse, whereas the \dj variant of \paco is consistently worse.}
\small
\tabcolsep 4pt
\centering\begin{tabular}{|c|ccc|ccc|ccc|ccc|ccc|}\hline
method & \multicolumn{3}{|c}{PACO} & \multicolumn{3}{|c|}{EPLL} & 
\multicolumn{3}{|c|}{BM3D} & \multicolumn{3}{|c|}{$\ell_1$}  & \multicolumn{3}{|c|}{DJ} 
\\
	percentile & \color{Gray} 25 & 50 & \color{Gray} 75 & \color{Gray} 25 & 50 & \color{Gray} 75 & \color{Gray} 25 & 50 & \color{Gray} 75& \color{Gray} 25 & 50 & \color{Gray} 75& \color{Gray} 25 & 50 & \color{Gray} 75 \\\hline
	$\sigma=10$ & 
	\color{Gray} 4.47 & \color{Black} 5.13 & \color{Gray} 6.06 & 
	\color{Gray} 4.48 & \color{Black} 5.12 & \color{Gray} 6.04 & 
	\color{Gray} 4.27 & \color{Black} 4.96 & \color{Gray} 5.79 & 
	\color{Gray} 4.47 & \color{Black} 5.15 & \color{Gray} 6.09 & 
    \color{Gray} 4.71 & \color{Black} 5.39 & \color{Gray} 6.21 \\ 
	$\sigma=20$ & 
	\color{Gray} 6.40 & \color{Black} 7.78 & \color{Gray} 9.18 & 
	\color{Gray} 6.44 & \color{Black} 7.83 & \color{Gray} 9.17 & 
	\color{Gray} 6.16 & \color{Black} 7.50 & \color{Gray} 8.98 &
	\color{Gray} 6.43 & \color{Black} 7.72 & \color{Gray} 9.12 &
	\color{Gray} 6.92 & \color{Black} 8.22 & \color{Gray} 9.97
	\\
	$\sigma=30$ & 
	\color{Gray} 8.25 & \color{Black} 9.88 & \color{Gray} 12.04 & 
	\color{Gray} 8.25 & \color{Black} 9.84 & \color{Gray} 11.87 & 
	\color{Gray} 7.50 & \color{Black} 9.10 & \color{Gray} 11.31 &
	\color{Gray} 8.25 & \color{Black} 9.88 & \color{Gray} 12.04 &
    \color{Gray} 8.69 & \color{Black} 10.67 & \color{Gray} 12.90	
	\\ \hline
\end{tabular}
\end{table}


\begin{figure}[p]
\centering\includegraphics[width=0.24\textwidth]{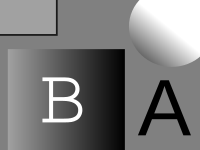} %
\includegraphics[width=0.24\textwidth]{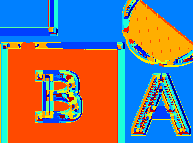} %
\includegraphics[width=0.24\textwidth]{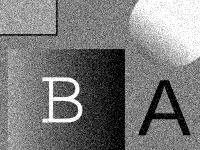} %
\includegraphics[width=0.24\textwidth]{fig/gmm_test/gt_modes.png}\\
\centering\includegraphics[width=0.24\textwidth]{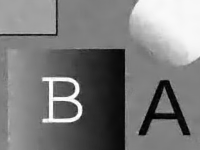} %
\includegraphics[width=0.24\textwidth]{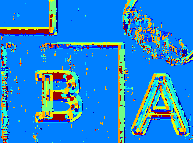} %
\includegraphics[width=0.24\textwidth]{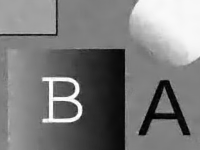} %
\includegraphics[width=0.24\textwidth]{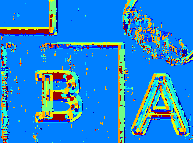}\\
\centering\includegraphics[width=0.24\textwidth]{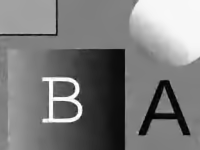} %
\includegraphics[width=0.24\textwidth]{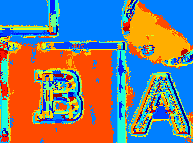} %
\includegraphics[width=0.24\textwidth]{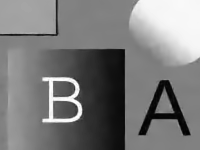} %
\includegraphics[width=0.24\textwidth]{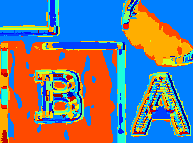}\\
\caption{\label{fig:denoising-results}%
Sample denoising results for test image and $\sigma=20$. Each pixel in the color images indicates which of the $200$ \gmm modes is assigned to each pixel in the corresponding image. The top row shows, from left to right: the clean image, its mode assignment map (this is the ground truth of the mode assignment problem), the noisy input, and the ground truth, again, for better comparison. The second row shows the first iterate of the denoised estimate obtained using \paco, its corresponding mode map, the first iterate of \epll, and its map. The last row shows the corresponding images for the 7th iterate of both algorithms (which is the last of \epll). As can be seen, the results are very close and both final maps are close to the ground truth. However, on close inspection, the \epll result has less artifacts and produces a softer cluster assignment. }
\end{figure}

\begin{figure}[p]
\includegraphics[width=0.24\textwidth]{fig/gmm_test/test_nacho.png}%
\includegraphics[width=0.24\textwidth]{fig/gmm_test/gt_modes.png} %
\includegraphics[width=0.24\textwidth]{fig/gmm_test/test_nacho_s20.png}%
\includegraphics[width=0.24\textwidth]{fig/gmm_test/gt_modes.png}\\
\includegraphics[width=0.24\textwidth]{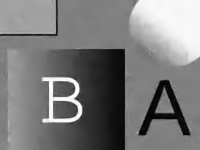}%
\includegraphics[width=0.24\textwidth]{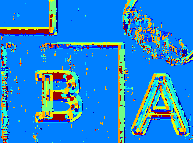} %
\includegraphics[width=0.24\textwidth]{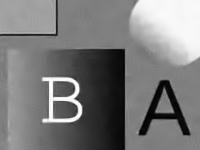}%
\includegraphics[width=0.24\textwidth]{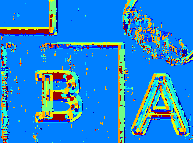}\\
\includegraphics[width=0.24\textwidth]{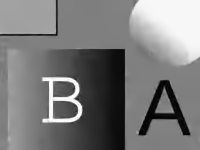}%
\includegraphics[width=0.24\textwidth]{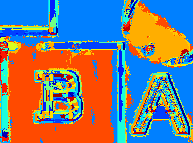} %
\includegraphics[width=0.24\textwidth]{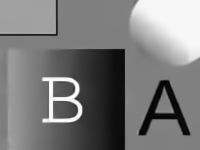}%
\includegraphics[width=0.24\textwidth]{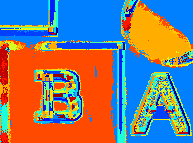}\\
\caption{\label{fig:denoising-results-variants}%
Comparison of denoising results obtained with the three \paco-\gmm variants on test image and $\sigma=20$. The first row is identical to that of \cref{fig:denoising-results}. The second and third rows correspond to the first and last iterates obtained using soft-thresholding ($\ell_1$) and \dj for the estimates instead of the baseline ($\ell_2$) estimator of \epll and  \paco-\gmm. As can be seen, the results are significantly better, both in perceptive quality, and in terms of the estimated mode assignment. }
\end{figure}

%
\begin{figure}[p]
\centering
\includegraphics[width=0.32\textwidth]{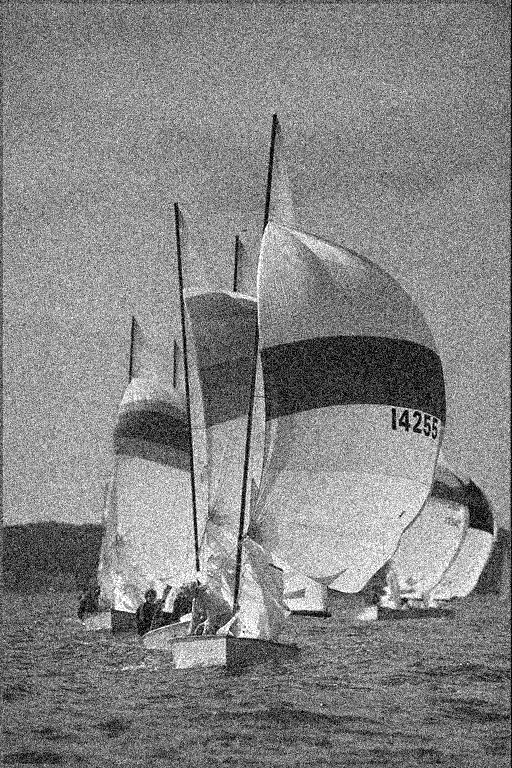}%
\includegraphics[width=0.32\textwidth]{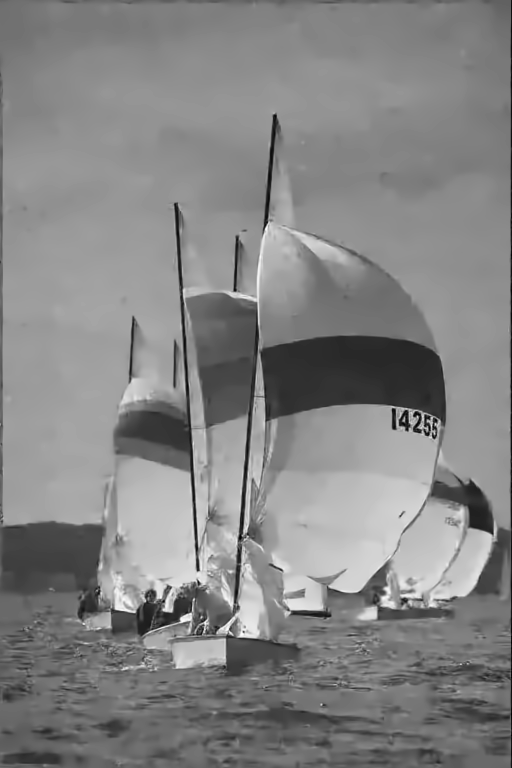}%
\includegraphics[width=0.32\textwidth]{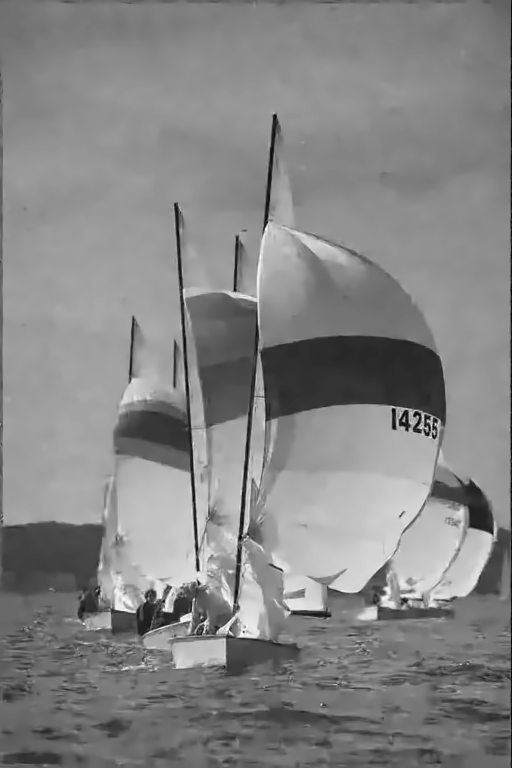}\\
\includegraphics[width=0.32\textwidth]{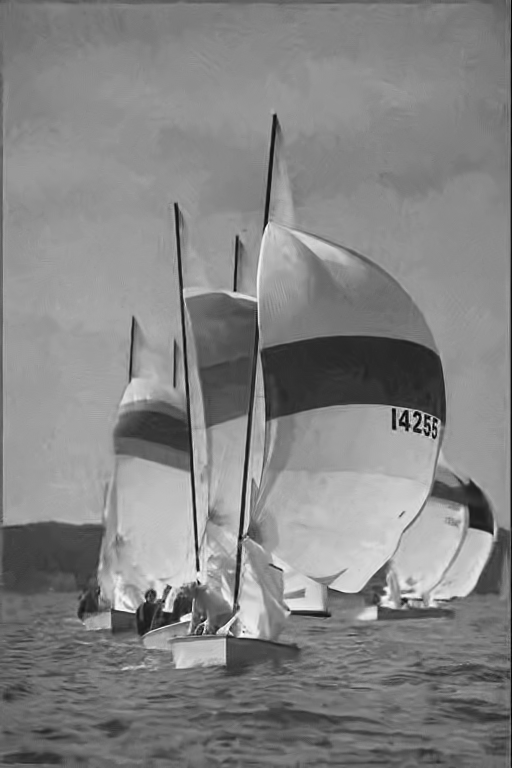}%
\includegraphics[width=0.32\textwidth]{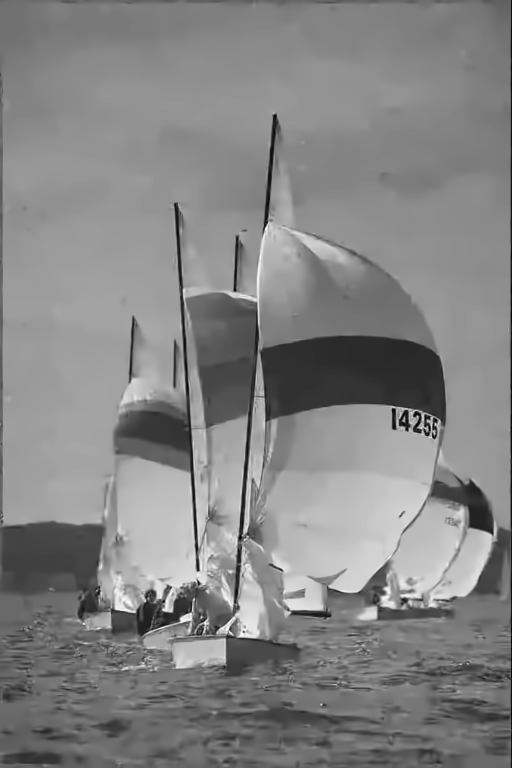}%
\includegraphics[width=0.32\textwidth]{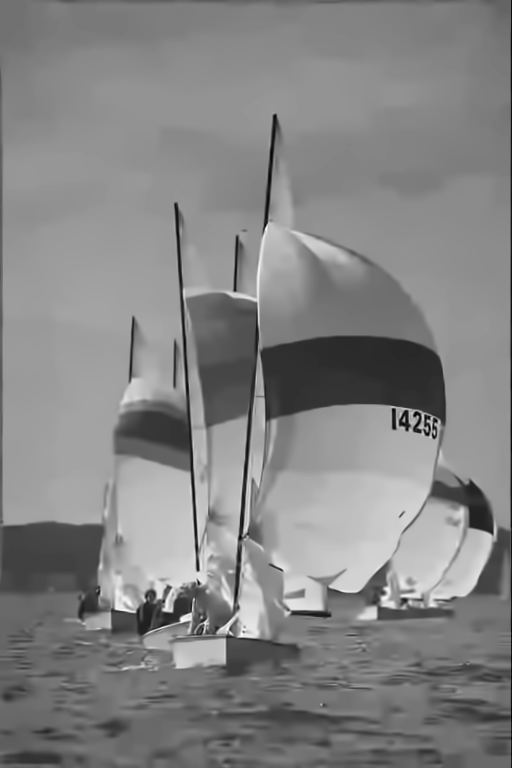}\\
\caption{\label{fig:denoising-results-kodak}%
Visual comparison of results on the Kodak dataset and all variants tested (high resolution image -- zoom for details). From top to bottom, left to right: noisy image (Kodim09 with $\sigma=30$), \paco-\gmm, \epll, BM3D, \paco-\gmm-$\ell_1$ and \paco-\gmm-\dj. The results in this case are clearly different between \epll and \paco-\gmm. Furthermore, despite their relatively low performance in terms of \rmse, the $\ell_1$ and \dj variants produce less artifacts and overall a nicer result than any of the other methods (BM3D, which consistently yields the lower \rmse, also exhibits more visible artifacts.)}
\end{figure}

\subsection{Quantitative comparison}

\Cref{tab:denoising-results} summarizes the results obtained on the Kodak dataset, under three levels of noise, by the three variants of \paco-\gmm described above, \epll, and BM3D. \Cref{fig:denoising-results-kodak} shows a visual comparison of these results on one of the Kodak images. Despite the similarities in terms of \rmse, the results  obtained with the different methods are clearly distinguishable. In particular, the $\ell_1$ and \dj variants exhibit less artifacts. Moreover, \textsc{BM3D}, which is better in terms of \rmse, tends to produce strong artifacts in smooth regions.

\section{Contrast Enhancement}
\label{sec:contrast}
\def\mone{\mathbf{1}}
\def\gce{\acro{gce}}
\def\lce{\acro{lce}}
\def\ace{\acro{ace}}
\def\cdf{\acro{cdf}}
\def\heq{\acro{heq}}

\begin{figure}
  \centering\hspace{4ex}\includegraphics[width=0.4\textwidth]{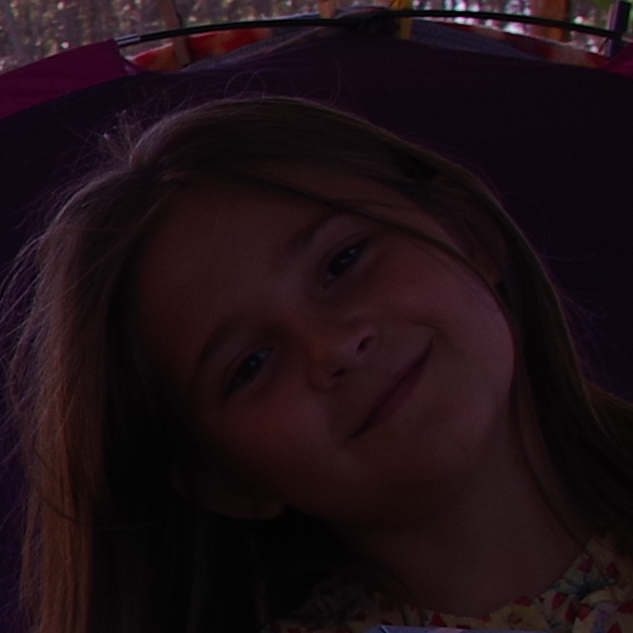}\hspace{7ex}%
  \includegraphics[width=0.4\textwidth]{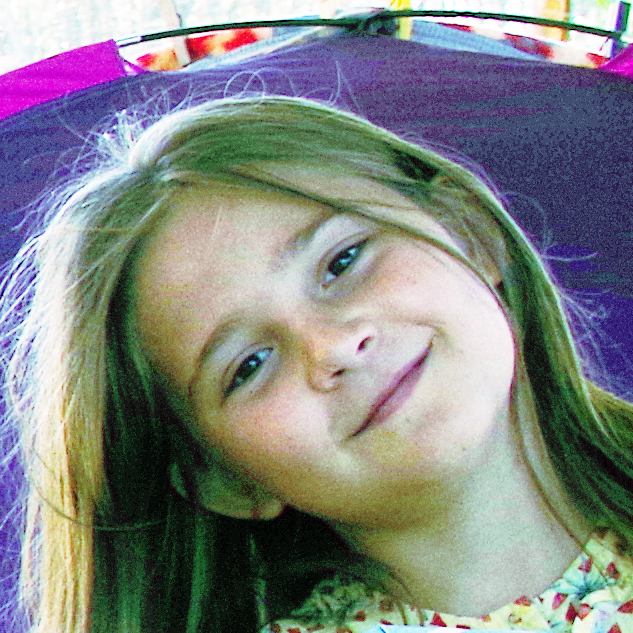}\\[1ex] 
  \centering\includegraphics[width=0.9\textwidth]{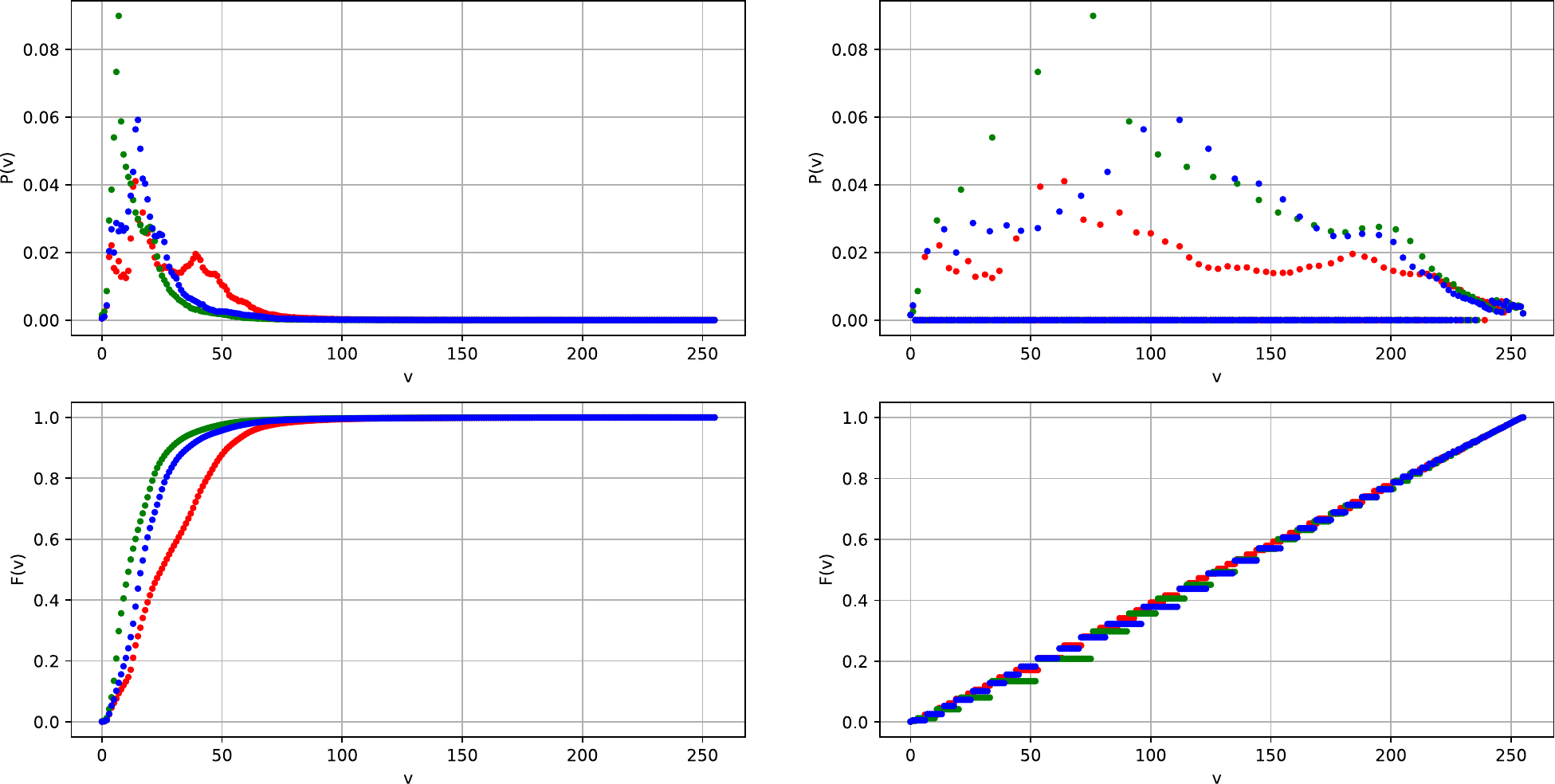}%
  \caption{\label{fig:global-heq} Global Histogram Equalization. Left, from top to bottom: original image, its histogram and \cdf. Right: enhanced image, its histogram, and its \cdf. In this case, the simple \gce is enough to produce a significantly better image.  }
  \end{figure}
  
As a final example application we explore a problem where the cost function is not derived from a prior probability model on the patches. This is the case of the Local Contrast Enhancement (\lce) problem: a now-ubiquitous post-processing image enhancement technique which generally provides better results than older and simpler Global Contrast Enhancement  (\gce) methods.

The basic method for improving the global contrast of an image is Histogram Equalization (\heq): given an input image $\vx$ taking on values in $[0,1)$, and the empirical cumulative distribution function of its values, 
$$F_{\vx}(z)=\frac{\sum_{i=1}^{N}\mone(x_i < z)}{N},$$
the equalized image $\hvx$ is obtained by the mapping 
\begin{equation*}
\hx_i \leftarrow F_{x}^{-1}(x_i). 
\end{equation*}
An example of this procedure is shown in \cref{fig:global-heq}.
 
\subsection{Contrast enhancement cost function}

Let $(0 \leq z_1 < z_2 < \ldots z_k \leq 1)$ be the $k$ different intensity levels occurring in $\vx$. Histogram equalization can be seen as finding  the corresponding levels in the enhanced image, $(\hz_1 < \ldots < \hz_k)$ which minimize the following cost function:
\begin{equation*}
h(\hvx) = \int_{0}^{1}{(F_{\hvx}(\mu)-\mu)}d\mu = \sum_{i=1}^{k+1}\int_{\hz_{i-1}}^{\hz_{i}}{(F_{\hvx}(\mu)-\mu)}d\mu,
\end{equation*}
where we have defined $z_0=0$ and $z_{k+1}=1$. Let $(p_1,p_2,\ldots,p_k)$ be the empirical probability distribution of each level in $\vx$; this vector is not modified during minimization. We have that 
$$F_{\hvx}(\mu)=F_{i-1}=\sum_{r=1}^{i-1}p_r,\;\hz_{i-1} \leq \mu < \hz_{i},
$$ 
so that $F_{\hvx}$ is constant within each integral and does not depend on $(\hz_1,\hz_2,\ldots,\hz_k)$. We define $F_0=0$. After integration, we obtain the following cost function on $(\hz_1,\hz_2,\ldots,\hz_k)$:
\begin{equation*}
h(\hvx) = f(\hz_1,\hz_2,\ldots,\hz_k)=\sum_{i=1}^{k+1}\int_{\hz_{i-1}}^{\hz_{i}}{(F_{\hvx}(\mu)-\mu)}d\mu = 
\sum_{i=1}^{k+1}
\frac{1}{2}
\left[
\left(F_{i-1}-\hz_i\right)^2-
\left(F_{i-1}-\hz_{i-1}\right)^2
\right].
\end{equation*}
The solution to the above problem is given by

\subsection{Local contrast enhancement}

In certain situations, global contrast enhancement is not enough. Images such as the leftmost one in \cref{fig:lce-results-1}, where large regions are subject to very different local luminosities, are not effectively improved by this method. As the name implies, local contrast enhancement works by correcting contrast in different regions. Two well known and very successful methods for this task are Retinex~\cite{retinex} and Automatic Color Enhancement (\ace)~\cite{ace}; see ~\cite{ace-ipol} for a fast, free and open source implementation of \ace. It should be noted that both Retinex and \ace are not limited to contrast enhancement: Retinex also handles white balance, while \ace handles contrast, white balance, and exposure automatically. The algorithm that we present next is limited only to contrast enhancement.
Besides these well established methods, there are many other local contrast enhancements  which have their own strengths, e.g. being faster, or being less sensitive to noise than the above ones~\cite{ms-retinex,piecewise-heq,lcc,variational-lce}. Many such methods are implemented in IPOL,\footnote{Image Processing On-Line, {http://ipol.im} an excellent resource for well documented and implemented image processing algorithms, which are free, open source, and can be experimented on-line.} see e.g.~\cite{variational-lce-ipol,ace-ipol,lcc-ipol,piecewise-heq,ms-retinex-ipol}; we invite the reader to experiment with these methods. 

In this section we explore the simple idea of applying histogram equalization on a patch-wise basis to achieve local contrast while using \paco to obtain a globally-coherent result. We do not claim our proposed method to be better than any of the aforementioned ones. In fact, it should hardly be the case, as many of them rely on carefully studied perceptual properties of human perception, whereas our method is conceptually very simple.

Because plain histogram equalization is usually very aggressive, we include an additional term to the cost function so that the result is a weighted average between the equalized result and the original patches. Also, in order to account include some exposure compensation capability (the so called \emph{gray world} principle), we add an extra term which accounts for deviations of the mean patch value from the middle intensity $0.5$; the resulting cost function on a given patch is given by:
$$
h_{\alpha,\beta}(\hvx_j) = \alpha\int_{0}^{1}{(F_{\hvx}(\mu)-\mu)}d\mu + 
\frac{\beta}{2}\left\|\hvx-\frac{1}{2}\mone\right\|_2^2 +
\frac{1-\alpha-\beta}{2}\left\|\vx-\hvx\right\|_2^2.
$$
The \emph{fidelity} term $\|\vx-\hvx\|_2^2$ can also be written in terms of the $F_i$ and $\hz_i$, so that we can write the cost function solely in terms of the output levels of the image patch:
\begin{equation}
h_{\alpha,\beta}(\hz_1,\hz_2,\ldots,\hz_k) = 
\frac{\alpha}{2}\sum_{i=1}^{k+1}
\left[
\left(F_{i-1}-\hz_i\right)^2-
\left(F_{i-1}-\hz_{i-1}\right)^2
\right] +
\frac{\beta}{2}\sum_{i=1}^{k}p_i(\hz_i-0.5)^2 +
\frac{1-\alpha}{2}\sum_{i=1}^{k}p_i(\hz_i-z_i)^2.
\label{eq:heq-mix}
\end{equation}
The solution to the above problem is 
\begin{equation}
\hvx = \alpha F^{-1}(\vx) + \beta\left(0.5-\frac{1}{N}\sum_i\vx_i\right)\mone + (1-\alpha-\beta)\vx.
\label{eq:heq-function}
\end{equation}

It is important to note that \cref{eq:heq-mix} includes a fidelity term with respect to the \emph{original} input image $\vx$. If we want to preserve its purpose, we cannot let this term depend on an image which is an iterate within the \admm algorithm, as the original reference would be lost. Thus, in our implementation, the last term of \cref{eq:heq-mix} is always with respect to the input image, which is constant throughout the iterations. With this in mind, we now plug $h_{\alpha,\beta}(\cdot)$ in the \paco framework to define our \paco-\heq problem:
\begin{equation*}
\hvx = \arg\min_{\zeta} h_{\alpha,\beta}(\zeta) + g(\zeta).
\end{equation*}
As in the denoising case, the splitting can be done in more than one way. We choose the following splitting, which is consistent with our definition of the problem:
\begin{equation*}
(\hvx,\hvz) = \arg\min_{\zeta,\xi} 
h_{\alpha,\beta}(\zeta)  + g(\xi) + 
\frac{1}{2\tau}\|\zeta-\xi\|_2^2,\;\st\;\zeta=\xi.
\end{equation*}
Following the standard procedure we have been using so far, the update $\hvy\iter{t+1}$ given $\hvz\iter{t}$ and $\vu\iter{t}$ is obtained by applying the stitching trick to $\hvz\iter{t}-\vu\iter{t}$. The update on $\hvz$ is given by 
$$
\hvz\iter{t+1} \leftarrow \frac{\alpha}{1+\tau}F^{-1}\left[\hvy\iter{t+1}-\vu\iter{t}\right]+
\frac{\beta}{1+\tau}\left[0.5-\frac{1}{N}\sum_i(\hvy\iter{t+1}-\vu\iter{t})_i\right]\mone +
\frac{1+\tau-\alpha-\beta}{1+\tau}\vx.
$$
Notice that the last term depends on the input $\vx$, not on the iterate. In the patch-wise formulation, this is replaced by the original image patches vector $\vy$. We summarize these steps in \cref{alg:paco-heq}.

\begin{algorithm}
\caption{\label{alg:paco-heq}PACO-HEQ algorithm. }
\begin{algorithmic}
\REQUIRE image $\vx$, mixing ratios $0 < \alpha,\beta < 1$, \admm parameter $\tau$
\STATE $\vy \leftarrow \mR\vx$
\STATE $t \leftarrow 0$; $\hvy\iter{t} \leftarrow \vy$; $\vu\iter{t} \leftarrow \mathbf{0}$
\REPEAT
\STATE  $\hvz_j\iter{t+1} \leftarrow  \frac{\alpha}{1+\tau}F^{-1}\left(\hvy_j\iter{t+1}-\vu_j\iter{t}\right)+
 \frac{\beta}{1+\tau}\left[0.5-\frac{1}{N}\sum_i(\hvy_j\iter{t+1}-\vu_j\iter{t})\right]\mone +
 \frac{1+\tau-\alpha-\beta}{1+\tau}\vy_j\,,\;j=1,\ldots,n$
\STATE  $\hvy\iter{t+1} \leftarrow \mR(\mR\transp\mR)\inv\mR\transp(\hvz\iter{t+1}+\vu\iter{t})$
\STATE  $\vu\iter{t+1} \leftarrow \vu\iter{t} + \hvy\iter{t+1} - \hvz\iter{t+1}$
\STATE  $t \leftarrow t+1$
\UNTIL{convergence}
\RETURN enhanced image $\hvx \gets \mR\transp\hvy\iter{t}$
\end{algorithmic}
\end{algorithm}

\subsection{Sample results}

\begin{figure}[htbp]
\centering%
\includegraphics[width=0.33\textwidth]{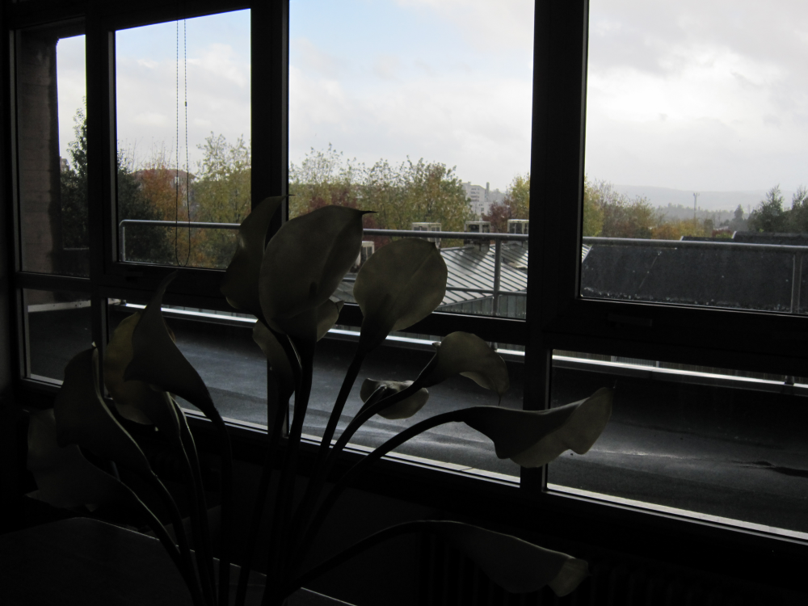}%
\includegraphics[width=0.33\textwidth]{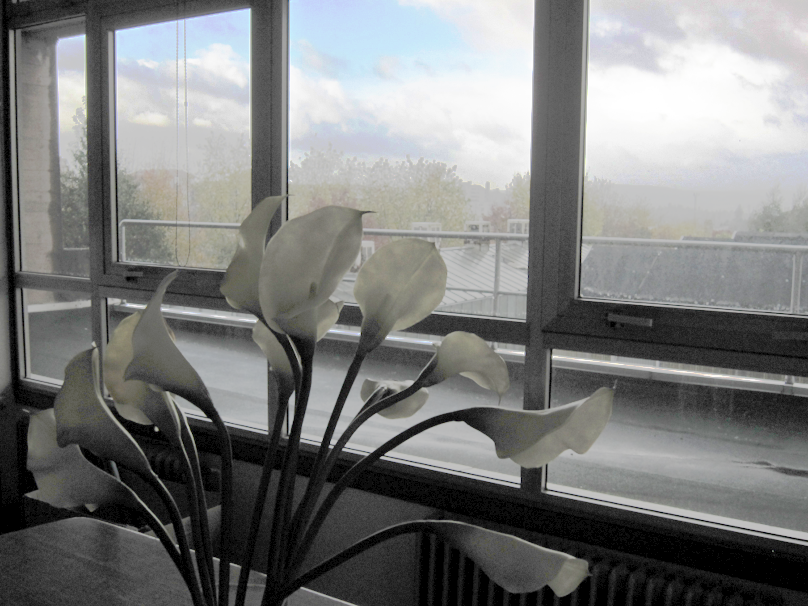}%
\includegraphics[width=0.33\textwidth]{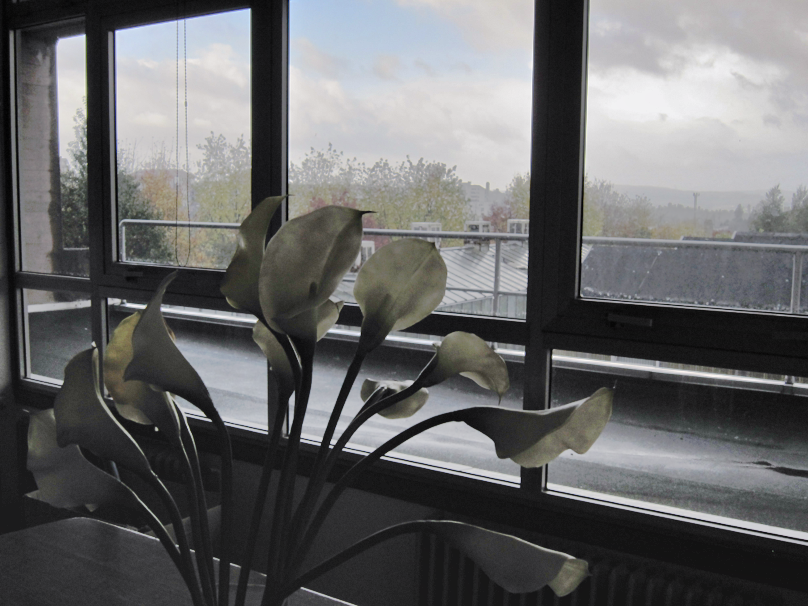}\\
\includegraphics[width=0.33\textwidth]{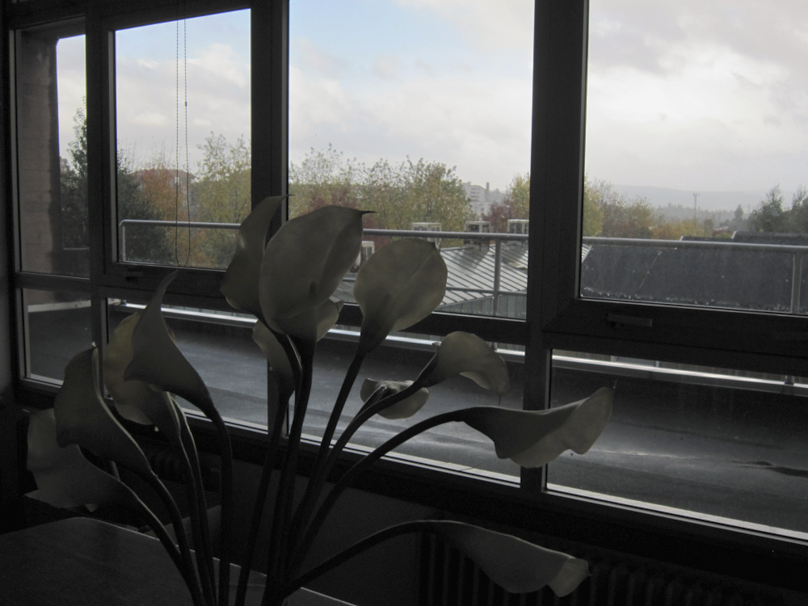}%
\centering%
\includegraphics[width=0.33\textwidth]{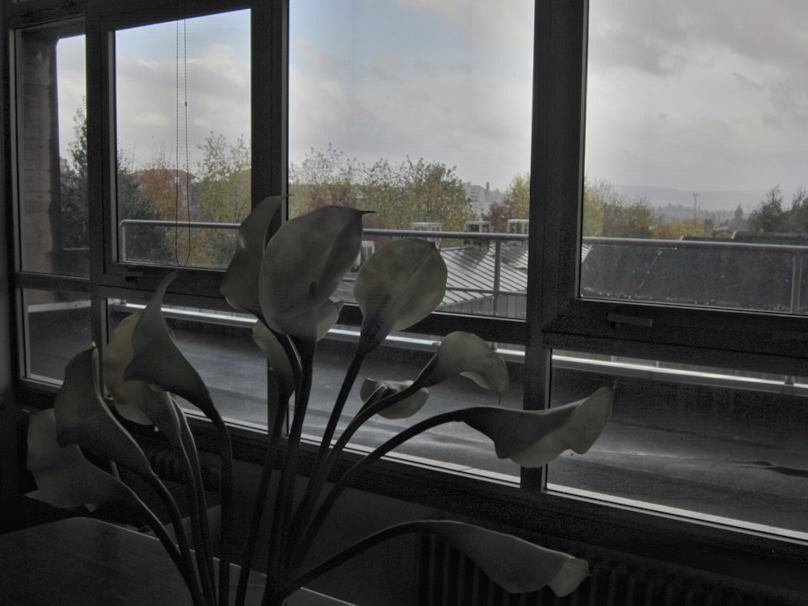}%
\includegraphics[width=0.33\textwidth]{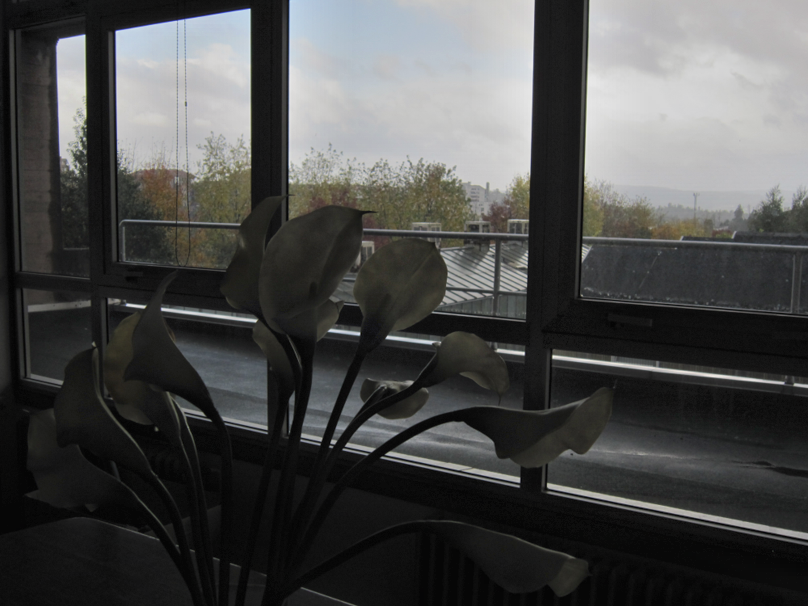}\\
\caption{\label{fig:lce-results-1} Contrast Enhancement results on Iris image. Top row: original image, global \heq, \ace with $\alpha=8$. Bottom row: \paco-\heq equalization function \cref{eq:heq-function} with $\alpha=0.25$ and $\beta=0.05$ applied globally, \paco-\heq with same parameters on patches of size $64{\times}64$ and stride $8$, \paco-\heq obtained with the same parameters except $\alpha=0.125$ (less contrast). Differences are particularly noticeable in the leaves of the plant. }
\end{figure}

\begin{figure}
\includegraphics[width=0.245\textwidth]{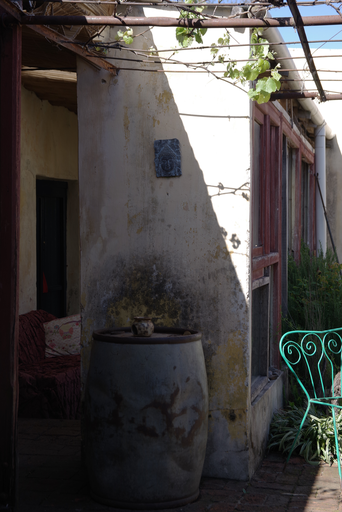}%
\includegraphics[width=0.245\textwidth]{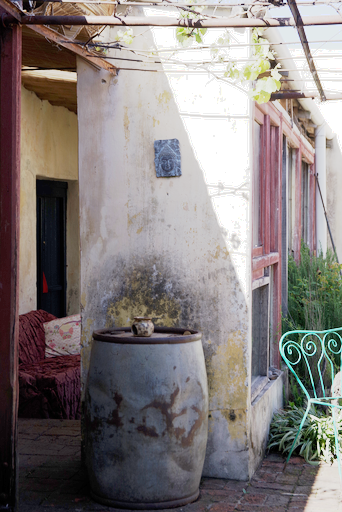}%
\includegraphics[width=0.245\textwidth]{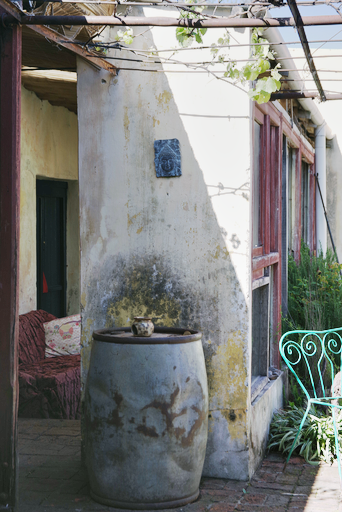}%
\includegraphics[width=0.245\textwidth]{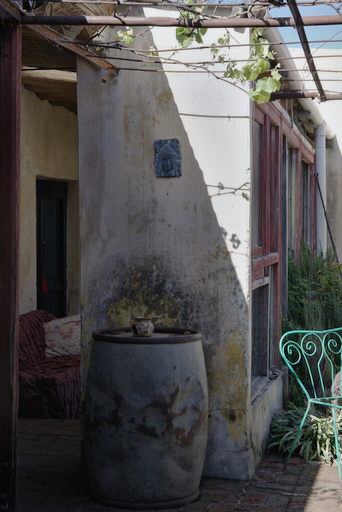}\\
\caption{\label{fig:lce-results-2} Contrast Enhancement results on Barril image. Left to right: original image, global HEQ, ACE with $\alpha=8$, \paco-\heq with $\alpha=0.25$,$\beta=0.05$, patches of size $64{\times}64$ and a stride of $s=8$. }
\end{figure}

Sample results are shown in \cref{fig:lce-results-1} and \cref{fig:lce-results-2}. Overall, with these parameters, \paco-\heq does produce a noticeable improvement in local contrast. It tends to preserve the local intensity more than the other methods, but still produces significantly sharper results. The source code, as well as more examples, can be found in~\url{http://iie.fing.edu.uy/~nacho/paco/}.

\section{Concluding remarks and future work}
\label{sec:conclusion}

We have presented \paco, a general framework for solving signal restoration problems under a patch consensus constraint. We have shown that the resulting optimization problem is feasible and easy to solve using the standard \admm method. In particular, we described a general and practical  solution to the patch reprojection step, named the ``stitching trick'', which sidesteps the major bottleneck in similar methods. We have also shown that the hard consensus constraint allows for the definition of  constraints directly in signal space, which may allow for new ways of formulating old problems.

The \paco framework was showcased in three sample applications with good results, both in terms of quality and efficiency. Several other applications are possible; we are already working on a few of them, the results of which will be published elsewhere. 

Besides new applications, \paco can also be used as a platform for distributed processing of very large images, by letting different nodes process smaller sub-images and then using the consensus mechanism to merge the results. In fact, distributed computing applications and network optimization are the fields where consensus optimization problems, as described in \cref{sec:paco}, were first analyzed.

Finally, the patch consensus idea can be extended in various ways. For example, it could be used in combination with the classic  \emph{mixture of experts} or \emph{boosting} as a way to combine the output of the experts in a meaningful way. 
Another interesting extension would be a true multiscale patch consensus framework (our current framework readily allows for schemes such as \emph{dilated convolutions}~\cite{dilated-conv}).


\bibliographystyle{siamplain}
\balance
\bibliography{IEEEabrv,paper,inpainting,contrast,denoising}
\end{document}